\documentclass[a4paper,11pt]{article}
\pdfoutput=1 % if your are submitting a pdflatex (i.e. if you have
\usepackage{jcappub}

\usepackage[T1]{fontenc} % if needed
\usepackage{placeins}
\usepackage[utf8]{inputenc}
\usepackage{amsmath,mleftright}
\usepackage{xparse}
\usepackage{listings}
\usepackage{caption}
\usepackage{subcaption}
\NewDocumentCommand{\evalat}{sO{\big}mm}{%
  \IfBooleanTF{#1}
   {\mleft. #3 \mright|_{#4}}
   {#3#2|_{#4}}%
}

\usepackage{graphicx}% Include figure files
\usepackage{dcolumn}% Align table columns on decimal point
\usepackage{amssymb}% outlined letters
\usepackage{amsmath}
\usepackage{color}
\usepackage{verbatim}
\usepackage{multirow}
\usepackage{soul}
\usepackage{parskip}
\usepackage{tabularx}
\usepackage[normalem]{ulem}  % for \sout (striking out)

\graphicspath{
{/Users/cg411/Documents/PhD/Projects/mcmc_2021/mcmc_runs/SNR_40/}
{./figures/spectral_params/}
{./figures/thermo_params/}
{./figures/induced_prior/}
{./figures/recon/}
}

\def\al{\alpha}
\def\be{\beta}

\def\ep{\epsilon}

\def\th{\theta}

\def\Om{\Omega}

\newcommand{\ben}{\begin{equation}}
\newcommand{\een}{\end{equation}}
\newcommand{\bea}{\begin{eqnarray}}
\newcommand{\eea}{\end{eqnarray}}
\newcommand{\ba}{\begin{array}}
\newcommand{\ea}{\end{array}}
\newcommand{\bit}{\begin{itemize}}
\newcommand{\eit}{\end{itemize}}

\newcommand{\A}{\mit{A}}
\newcommand{\E}{\mit{E}}
\newcommand{\T}{\mit{T}}

\newcommand{\cs}{c_\text{s}} % Sound speed

\newcommand{\Dbin}{\bar{D}_{b}}

\newcommand{\DAbin}{\bar{D}^{\A}_b}
\newcommand{\DEbin}{\bar{D}^{\E}_b}

\newcommand{\fp}{f_\text{p}}

\newcommand{\Fgwt}{F_{\text{gw,0}}}

\newcommand{\HN}{H_\text{n}} % Hubble rate at nucelation
\newcommand{\Hn}{H_\text{n}}

\newcommand{\NA}{N_{\A}}
\newcommand{\NE}{N_{\E}}

\newcommand{\nbin}{n_{b}}
\newcommand{\Nbin}{N_\text{b}}
\newcommand{\Nacc}{N_\text{acc}}
\newcommand{\Npos}{N_\text{oms}}

\newcommand{\OmGW}{\Omega_\text{gw}}
\newcommand{\OmGWSSM}{\Omega_\text{gw}^\text{ssm}}

\newcommand{\OmPeak}{\Omega_\text{p}}

\newcommand{\OmGWtSSM}{\Omega_{\text{gw,0}}^{\text{ssm}}}
\newcommand{\OmGWtdbp}{\Omega_{\text{gw,0}}^{\text{dbp}}}

\newcommand{\OmInst}{\Omega_{\mathrm{ins}}}% data gw signal

\newcommand{\PspecGWhat}{\hat{\mathcal{P}}_{\text{gw}}}
\newcommand{\Pacc}{P_{\text{acc}}}
\newcommand{\Poms}{P_{\text{oms}}}

\newcommand{\rb}{r_\text{b}} % Break frequency ratio

\newcommand{\SpecDenGW}{\tilde P_{\text{gw}}}

\newcommand{\SNR}{\rho}

\newcommand{\TN}{T_\text{n}} % Nucleation temperature
\newcommand{\Tn}{T_\text{n}} % Nucleation temperature
\newcommand{\Tobs}{T_{\text{obs}}}

\newcommand{\vw}{v_\text{w}} % Wall speed
\newcommand{\spectralparamswprior}{(\log_{10} \OmPeak, \log_{10}  (\fp /{\rm Hz}), \rb , b)}
\newcommand{\thermoparamswprior}{(\log_{10}(\Tn/\mathrm{Gev}),\log_{10} \al, \log_{10} r_* , \vw)}

\definecolor{darkgreen}{rgb}{0.0,0.5,0.0}

\definecolor{teal}{rgb}{0.25, 0.4, 0.96}

\title{\LARGE Reconstructing physical parameters from template gravitational wave spectra at LISA: first order phase transitions}
\subheader{HIP-2022-23/TH, ULB-TH/22-13}
\author[a,b]{Chloe Gowling,}
\author[b,a]{Mark Hindmarsh,}
\author[b]{Deanna C. Hooper,}
\author[c]{Jes\'us Torrado}

\affiliation[a]{Department of Physics and Astronomy, University of Sussex, BN1 9QH, Brighton, UK}
\affiliation[b]{Department of Physics and Helsinki Institute of Physics, PL 64, FI-00014 University of Helsinki, Finland}
\affiliation[c]{Service de Physique Th\'eorique, Universit\'e Libre de Bruxelles, Boulevard du Triomphe CP225, B-10503 Brussels, Belgium}

\emailAdd{c.gowling@sussex.ac.uk}
\emailAdd{mark.hindmarsh@helsinki.fi}
\emailAdd{deanna.hooper@helsinki.fi}
\emailAdd{jesus.torrado.cacho@ulb.be}

\abstract{A gravitational wave background from a first order phase transition in the early universe may be observable at millihertz gravitational wave (GW) detectors such as the Laser Interferometer Space Antenna (LISA). In this paper we introduce and test a method for investigating LISA's sensitivity to gravitational waves from a first order phase transition using parametrised templates as an approximation to a more complete physical model.  The motivation for developing the method is to provide a less computationally intensive way to perform Markov Chain Monte Carlo (MCMC) inference on the thermodynamic parameters of a first order phase transition, or on generally computationally intensive models. Starting from a map between the physical parameters and the parameters of an empirical template, we first construct a prior on the empirical parameters that contains the necessary information about the physical parameters; we then use the inverse mapping to reconstruct  approximate posteriors on the physical parameters from a fast MCMC on the empirical template.
We test the method on a double broken power law  approximation to spectra in the sound shell model. The reconstruction method substantially reduces the proposal evaluation time,  and despite requiring some precomputing of the mapping, this method is still cost-effective overall. In two test cases, with signal-to-noise $\sim 40$, the method recovers the physical parameters and the spectrum of the injected gravitational wave power spectrum  to $95\%$ confidence.
In previous Fisher matrix analysis we found the phase boundary speed $\vw$ was expected to be the best constrained of the thermodynamic parameters. In this work,  for an injected  phase transition GW power spectrum with   $\vw = 0.55$, with a direct sample on the thermodynamic parameters we recover $0.630^{+0.17}_{-0.059}$ and for our reconstructed sample $0.646^{+0.098}_{-0.075}$. 
}

\begin{document}
\maketitle
\flushbottom

\setlength{\parskip}{6pt plus 1pt}

\section{Introduction}

The Laser Interferometer Space Antenna (LISA), due to launch in the 2030s \cite{Audley:2017drz} will probe the previously unexplored  millihertz region of the gravitational wave (GW) spectrum.  The LISA sensitivity window,  $10^{-4}$ Hz to $10^{-1}$ Hz, has an abundance of GW sources ranging from astrophysical:  black hole mergers, galactic binaries \cite{Postnov:2014tza}, extreme mass ratio binaries \cite{AmaroSeoane:2007aw} and precursors for stellar origin black hole mergers \cite{Sesana:2016ljz};
to the cosmological: cosmic strings, inflation and phase transitions ~\cite{Caprini:2018mtu,LISACosmologyWorkingGroup:2022jok}. Here we focus on LISA's sensitivity to the cosmological  stochastic GW background (SGWB) from a first order phase transition.

The early universe was hot and dense; as it expanded and cooled the universe may have undergone several phase transitions. In particular, we are interested in a possible first order phase transition associated with electroweak symmetry breaking. In the Standard Model this process occurs via a crossover and no GWs are produced \cite{Kajantie:1996qd,Kajantie:1996mn}. Alternatively, in numerous extensions to the Standard Model, in some cases motivated as explanations for dark matter or the baryon asymmetry of the universe, a first order phase transition and the production of GWs is possible. See \cite{Caprini:2019egz} for a review of models.

In a first order phase transition, once a critical temperature is reached, bubbles of the broken phase nucleate in the symmetric phase; these bubbles expand, collide, and percolate until the phase transition is complete. During this process GWs are produced via the collisions of bubble wall, the subsequently produced sound waves and turbulent flows. For a review of first order phase transitions see \cite{Mazumdar:2018dfl,Hindmarsh:2020hop,Weir:2017wfa}.

If the first order phase transition is driven by thermal fluctuations, the acoustic source of GWs dominates \cite{Hindmarsh:2013xza,Hindmarsh:2015qta,Hindmarsh:2017gnf}; production by bubble collisions \cite{Cutting:2018tjt,Cutting:2020nla,Lewicki:2020azd,Lewicki:2020jiv,Lee:2021nwg,Lewicki:2022pdb} can become relevant if there is very strong supercooling \cite{Ellis:2019oqb,Ellis:2020nnr}. 
Here we assume that the sound wave component is dominant, and model the GW background component with the Sound Shell Model (SSM)~\cite{Hindmarsh:2016lnk, Hindmarsh:2019phv}. The SGWB from a first order phase transition is determined by key thermodynamic parameters: the nucleation temperature $\Tn$, the phase transition strength $\al$, the wall speed $\vw$ and the Hubble-scaled mean bubble spacing $r_*$. The speed of sound, which can take different values in the two phases, also impacts the SGWB produced \cite{Giese:2020znk, Giese:2020rtr}; however for this first analysis we take it to be the ultra-relativistic value of  $1/\sqrt{3}$.

Calculating numerous GW power spectra for a first order phase transition using the SSM, as when one conducts Markov Chain Monte Carlo (MCMC) analyses, is computationally expensive. This motivates the use of a fit to the phase transition SGWB that is quick to evaluate: here we use a double broken power law, which provides a good fit to the SSM over most of the thermodynamic parameter space \cite{Gowling:2021gcy, Hindmarsh:2016lnk}. The double broken power law is characterised by four ``spectral'' parameters: the peak amplitude $\OmPeak$, the peak frequency $\fp$, the ratio $\rb$ between the peak frequency and the break frequency, and the slope between the two characteristic frequency scales $b$. The ultimate goal is to infer the thermodynamic parameters of a supposed SGWB signal by fitting to it a computationally cheap double broken power law.

To achieve this, we require a robust method for transforming information about the spectral parameters into constraints on the thermodynamic ones. The first step is to transform the physically-motivated prior density on the thermodynamic parameters into an induced prior on the spectral ones, which is achieved by weighting an initial spectral prior with the density of the image of a prior-consistent grid of thermodynamic parameters in the spectral parameter space. Constraints on the spectral parameters obtained with such a prior can then be translated back to the thermodynamic parameter space by using the inverse of the projection that we just described. This reconstruction method is a general cost-effective preliminary parameter estimation framework that can be applied to any model for which computing the SGWB is expensive, but for which there exists a reasonably good empirical approximation.

As a demonstration, in this study we consider two fiducial models with different thermodynamic parameters, and use MCMC methods to estimate LISA's ability to perform parameter estimation for both the spectral and the (much slower) thermodynamic parameterisations. We then compare the latter result with the constraints on the thermodynamic parameters derived from the spectral parameter sample using our reconstruction methodology. We consider a data model made of the phase transition SGWB and LISA noise. In a global fit the impact of astrophysical foregrounds from the extragalactic black holes, binary neutron stars and double white dwarf populations should also be considered; in this first investigation of the method we ignore these foregrounds. However, their impact on the MCMC estimation of spectral parameters has recently been considered in \cite{spec_separation_2022}. Parameterised templates with more general spectral forms have been explored in \cite{Giese:2021dnw}; although no reconstruction of the underlying parameters was attempted.

This paper is structured as follows: in Section \ref{sec:sgwbpt} we describe the expected SGWB spectrum from cosmological first order phase transitions in the SSM; in Section \ref{sec:noise} we describe the LISA noise model, and in Section \ref{sec:datainference} we go on presenting our data model, and the likelihood and base priors that will be used; in Section \ref{Sec:recon} we present the reconstruction algorithm, and finally in Section \ref{sec:results} we apply it to the aforementioned fiducial models. We lay out our conclusions and discuss some future prospects in Section \ref{sec:conclusions}.

\section{SGWB from cosmological first order phase transition}
\label{sec:sgwbpt}

The GW power spectrum from a first order phase transition can be characterised by the thermodynamic parameters %($\vw,\al, r_*,\Tn $).
($\Tn, r_*, \al, \vw$). 
 Firstly, the nucleation temperature $\Tn$, is the temperature corresponding to the peak of the globally-averaged bubble nucleation rate.  The Hubble rate at the nucleation temperature $\Hn$ sets the frequency scale of the GW spectrum.

The second thermodynamic parameter is the nucleation rate parameter $\be$. As discussed in \cite{Gowling:2021gcy}, due to uncertainties in the calculation of $\be$, we instead consider the related quantity, the mean bubble spacing $R_*$. We note that $\be^{-1}$ is the time for the bubble wall to move a distance $R_*$ and therefore has the interpretation of the duration of the phase transition. In this work we refer to the Hubble-scaled mean bubble spacing $r_* = \Hn R_*$ which contributes to  the frequency scale and amplitude of the GW power spectrum.

Our third key thermodynamic parameter is the phase transition strength $\al$, which we define as the ratio between the trace anomaly and the thermal energy, where the trace anomaly describes the amount of energy available for conversion to shear stress energy. A stronger transition means more energy is converted to shear stress energy and a larger overall amplitude for the GW signal.

The final parameter to introduce is the wall speed $\vw$ which, along with $\al$, determines the motion of the plasma surrounding the bubble wall. The value of the wall speed relative to the speed of sound $\cs$ determines the width of the GW power spectrum, here we assume the ultrarelativistic value $\cs = 1/\sqrt{3}$ (see \cite{Giese:2020rtr,Giese:2020znk} for other scenarios). For wall speeds close to $\cs$ the power spectra are broad and $\rb$ (the ratio between the peak frequency and the break frequency) is small, in the opposite case the power spectra are narrow.

The general form of the gravitational wave power spectrum from a thermal first order phase transition is
 \ben\label{eq:Omgw_ssm}
 \OmGW(z) = 3K^{2}(\vw,\al)\left(\HN \tau_{\mathrm{v}}\right)\left(\HN R_*\right) \frac{z^{3}}{2 \pi^{2}} \SpecDenGW\left(z\right),
 \een
 where $R_*$ is the mean bubble spacing, $z = k R_*$, $k$ is the comoving wavenumber and $K(\vw,\al)$ is the fraction of the total energy converted into kinetic energy of the fluid.  The Hubble rate at nucleation is $\HN$, $\tau_v $ is the lifetime of the shear stress source,
the factor $R_*$ appears as an estimate of the source coherence time and $\SpecDenGW \left(z\right)$ is the dimensionless shape spectral density.
Eq.~\eqref{eq:Omgw_ssm} can be regarded as the definition of $\SpecDenGW$. As introduced and discussed in \cite{Gowling:2021gcy}, for simplicity we define
\ben\label{eq:scaling_factors_GW}
J = \Hn R_* \Hn \tau_v  = r_* \left(1 -  \frac{1}{\sqrt{1 + 2x}} \right).
\een
where 
%$r_* = \HN R_*$,  which we refer to as the Hubble-scaled mean bubble spacing, and 
$x = \HN R_* / \sqrt{K} $ is the ratio of the Hubble time $\HN^{-1}$ and the fluid shock appearance time $\tau_\text{sh} = R_* / \sqrt{K}$  \cite{Dahl:2021wyk}. 
The second equality is a model for the lifetime of the shear stress source in an expanding universe \cite{Guo:2020grp}.

\subsection{Gravitational wave power spectrum in the SSM}
Here we focus on the contribution from the sound waves and use the Sound Shell Model \cite{Hindmarsh:2019phv, Hindmarsh:2016lnk}, which limits us to transitions which are not so strong that the modifications to the spectrum from shocks \cite{Dahl:2021wyk} and vortical turbulence \cite{Auclair:2022jod} become important.
We use the \texttt{PTtools}\footnote{Code available on request to MH.} module which uses the SSM to directly compute the scale-free gravitational wave power spectrum $\PspecGWhat$ for a given $\vw$  and $\al$ \cite{Hindmarsh:2019phv}, 
defined as
 \ben{\label{eq:power_gw_scaled_output}}
\PspecGWhat (z)= 3K^{2} \frac{z^{3}}{2 \pi^{2}} \SpecDenGW \left(z\right).
\een
 The specifications of the calculations done with \texttt{PTtools} are the same as used in our previous work \cite{Gowling:2021gcy}.
We now introduce
\ben\label{eq:omgw_ssm}
 \OmGWSSM(z) =  J \PspecGWhat (z).
\een
As discussed in \cite{Gowling:2021gcy},  recent 3d-hydro simulations for $\al$ up to $\mathcal{O}$(1) (strong transitions) found that as transition strength increases, the efficiency of fluid kinetic energy production is less than previously expected \cite{Cutting:2019zws}. We estimate suppression in gravitational waver power observed in the numerical simulations, as a factor $\Sigma(\vw,\al)$.  For a complete outline of how we calculate $\Sigma $ see Appendix A in \cite{Gowling:2021gcy}.
The gravitational wave power spectrum at dimensionless comoving wavenumber $z$ just after the transition, and before any further entropy production, is then
\ben\label{eq:SSM_suppressed}
\Om_{\text{gw}}(z) = \OmGWSSM(z)\Sigma(\vw,\al).
\een
Today the power spectrum at physical frequency $f$ is
\ben{\label{eq:Omgw0_sup}}
    \OmGWtSSM(f) =\Fgwt \Om_{\text{gw}}(z(f)),
\een
where
\ben{\label{eq:Fgw0_def}}
\Fgwt=\Omega_{\gamma, 0}\left(\frac{g_{s 0}}{g_{s *}}\right)^{\frac{4}{9}} \frac{g_{*}}{g_{0}} = (3.57 \pm 0.05) \times 10^{-5} {\bigg( \frac{100}{g_*}\bigg)}^{\frac{1}{3}}
\een
is the power attenuation following the end of the radiation era.
Here $\Omega_{\gamma, 0}$ is the photon energy density parameter today, $g_{s }$ denotes entropic degrees of freedom and $g$ describes the pressure degrees of freedom.
In both cases the subscripts $0$ and  $*$ refer to their value today and the value at the time the GWs were produced respectively.
We evaluate $\Fgwt$ with the values given in \cite{Caprini:2019egz}, and use a reference value $g_* = 100 $.

We convert from dimensionless wavenumber $z$ to frequency today by taking into account redshift
\ben
\label{e:fzrstar}
f =\frac{z }{r_*} f_{*,0},
\een
where
\ben {\label{eq:f0} }
f_{*,0}=  2.6 \times 10^{-6} \,\textrm{Hz} \left(\frac{\TN}{100\,\textrm{GeV}}\right)\left(\frac{g_*}{100}\right)^{\frac{1}{6}},
\een
is the Hubble rate at the phase transition redshifted to today \cite{Caprini:2019egz}.
We assume the phase transition takes place well within one Hubble time so all frequencies throughout the transition have the same redshift.

\subsection{Double broken power law}
In the SSM there are two characteristic length scales, the mean bubble separation and the sound shell thickness, which motivate a simplified description in terms of a function with two frequency scales and three power law indices - a double broken power law \cite{Hindmarsh:2019phv}. The power spectrum today for the double broken power law fit can be described as
\ben{\label{eq:omgw_dbl_brkn}}
    \OmGWtdbp(f,\OmPeak,\fp,\rb, b) =\OmPeak M(s,\rb, b)
\een
where $\OmPeak$ is the peak of the power spectrum, $s = f/\fp$, $\fp$ is the frequency corresponding to $\OmPeak$ and  $\rb =  f_{\text{b}} /\fp$ is the ratio between the two breaks in the spectrum. The parameter $b$ defines the spectral slope between the two breaks. The spectral shape $M(s,\rb, b)$ is a double broken power law with a spectral slope $9$ at low frequencies and $-4$ at high frequencies, a form that was chosen to best describe the SSM \cite{Hindmarsh:2019phv}.

\ben{\label{eq: M double_break}}
    M ( s, \rb , b ) = s^ { 9 } {\left( \frac { 1 + \rb^4 } { \rb^4 + s^4}\right)}^{(9 -b)/4}  \left( \frac { b +4 } { b + 4 - m + m s ^ { 2 } } \right) ^ { (b +4) / 2 } .
\een
Within $ M ( s, \rb , b )$, $m$ has been chosen to ensure that  for $\rb<1$ the peak occurs at $s=1$  and $M(1,\rb,b) = 1$, giving
\ben{\label{eq: m}}
    m = \left( 9 {\rb}^4+ b\right) / \left( {\rb}^4 +1 \right).
\een

\section{LISA instrument noise model}
\label{sec:noise}

LISA will be a triangular constellation of three spacecraft connected via lasers with arm length of $2.5$ million km. Passing GWs will induce a distance modulation in the instrument arm length that is measured via the phase differences between lasers on the local and remote spacecraft. The phase differences (interferometer signals) can be combined in different ways with different time delays to eliminate the laser noise  \cite{Tinto:2001ii,Tinto:2002de}. We follow the convention for the three noise-orthogonal time delay interferometry (TDI) variables $A$, $E$ and $T$, as described in \cite{Smith:2019wny}. The $\T$ variable can be approximated as being insensitive to GWs. Here we assume the instrument noise is completely known and build our data model combining the $\A$ and $\E$ channels.

We construct the instrument power spectral density following the conventions given in \cite{Smith:2019wny} and used in  \cite{Gowling:2021gcy}. For the LISA instrument noise model we use the functions and parameter values given in the LISA Science Requirements Document \cite{LISA_SR_doc}.  In the $A$ and $E$ TDI channels the instrument noise spectral density 
arising from the optical metrology system noise (oms) and the test mass acceleration noise (acc)
is given by
\ben
\NA = \NE  = N_1 - N_2 \simeq (6 \Poms + 24 \Pacc) |W(f)|^2,
\een
where
\bea
N_1 &=& [4\Poms (f) + 8 \left[ 1 + \cos ^2 (f/f_*) \right] \Pacc(f) ] |W(f)|^2 , \\
N_2 &=& - [\Poms(f)  + 8 \Pacc] \cos(f/f_*) |W(f)|^2,
\eea
and $W(f) = 1 - \exp( 2 if/f_*)$, representing the interference induced by a return journey along one arm.  
In the above $f_*= c/(2 \pi L)$ is the transfer frequency, $L = 2.5\times10^9\;\text{m}$ is the constellation arm length, $c$ is the speed of light, 
and the model for the noise is  
\bea
\Poms &=& \Npos, \\
\Pacc &=&  \frac{ \Nacc}{(2\pi f)^{4}}  \left(  1 + \left(\frac{f_1}{f}\right)^2\right),
\eea
with $\Nacc = 1.44 \times 10^{-48} \; \text{s}^{-4} \text{Hz}^{-1}$, $\Npos =3.6 \times 10^{-41}\;\text{Hz}^{-1}$
and $f_1 = 0.4$ mHz \cite{LISA_SR_doc}.

To take into account the detector response to incident GWs, we consider the sensitivity $S$  for the $\A$ and $\E$ channels,
\ben\label{eq:SA}
S_{{A}}=  S_{{E}} =\frac{\NA}{\mathcal{R}_{\A}} \simeq \frac{40}{3}\left(\Poms + 4\Pacc \right)\left[1+\left(\frac{f}{4 f_*/ 3}\right)^{2}\right],
\een
where $\mathcal{R}$ is the detector response to isotropic stochastic GWs. In general, $\mathcal{R}$ must be evaluated numerically;  
here we use the simpler analytic fits presented in \cite{Smith:2019wny}
\ben\label{eq:R_A_fit}
\mathcal{R}^{\rm Fit}_{A}(f) = \mathcal{R}^{\rm Fit}_{E}(f) =\frac{9}{20} |W(f)|^2 \left[1 + \left(\frac{ f}{4f_*/3}\right)^2 \right]^{-1}.
\een
The sensitivities can be thought of as GW signals with unit signal-to-noise ratio at all frequencies.

In this work we will be interested in the sensitivity expressed as a GW fractional energy density power spectrum, related to the sensitivity by
\ben{\label{eq:OmInst}}
\OmInst =\left(\frac{4 \pi^{2}}{3 H_{0}^{2}}\right) f^{3} S_{\A}(f) ,
\een
which we will refer to as the LISA instrument noise.
The fiducial models have a signal-to-noise ratio  $\rho$ of approximately $\rho \approx 40$.  As we will show in the next section, our data model will combine the $\A$ and $\E$ channels, and the corresponding signal-to-noise ratio is given by  \cite{Smith:2019wny}
\ben\label{eq:SNR}
\rho =\sqrt{ 2\Tobs \int^\infty_0 df \frac{\Omega_{\text{gw}}^2}{\OmInst^2}} \ .
\een
As the $T$ channel is insensitive to GW signatures at low frequencies, it allows the instrument noise  at low frequencies to be better characterised.

\section{Parameter inference from mock LISA data}
\label{sec:datainference}

In this section we describe the data model used for LISA, the likelihood used for parameter inference from an injected SGWB, and priors for both thermodynamic and spectral parameters.

\subsection{Data model and likelihood }
\label{sec:datalike}
Here we outline how we model the LISA data, explain the assumptions made, and define the
likelihood used. The LISA data is expected to be a $\Tobs$ = 4 yr stream with a regular data
sampling interval $\T_{\text{samp}}$ = 5 s, not taking into account scheduled maintenance breaks. We use the data model as described in our previous work \cite{Gowling:2021gcy}.

In this analysis we consider the $\A$ and $\E$ TDI channels in the frequency domain, 
binned into $\Nbin = 1000$ logarithmically spaced positive frequency bins, with power
spectral densities $\DAbin$ ,$\DEbin$. The variance of the $\A$ and $\E$ channels are taken to be independent and identical. Within each bin there are $\nbin$ frequencies
\ben\label{eq:nbin}
\nbin = [(f_b- f_{b-1})\Tobs]
\een
where the square brackets denote the integer part, and here $\nbin \gg 1$, which justifies the use of a Gaussian likelihood. 
We combine the $A$ and $E$ data channels $\Dbin = (\DAbin + \DEbin )/2$, so that the log-likelihood for the spectral parameter case is then given by 
\ben\label{eq:log_like}
l = -\frac{1}{2} \sum_{b=1}^{\Nbin} \frac{2\nbin\left( \Omega_{\textrm{t}}(f_b,\theta) - \Omega_{\textrm{fid}}(f_b,\tilde{\th}_{\rm fid})\right)^{2}}{ \Omega_{\textrm{t}}(f_b,\theta)^2},
\een
where $ \Omega_{\textrm{fid}}, \Omega_{\textrm{t}}$ are related to the power spectral densities as described in Eq.~\ref{eq:OmInst} and $\tilde{\th}_{\rm fid}$ describes the fiducial model. The theoretical model of the data is given by
\ben{\label{eq:Omn}}
\Omega_{\textrm{t}}(f_b,{\theta}) = \OmInst(f_b) + \Om_{\textrm{pt}}(f_b,\th),
\een
where $\Om_{\textrm{pt}}(f_b,\th)$ is described by Eq.~\eqref{eq:omgw_dbl_brkn}. The thermodynamic case is obtained by replacing  $\Om_{\textrm{pt}}(f_b,\th)$ with $\Om_{\textrm{pt}}(f_b,\tilde{\th})$ which is described by Eq.~\eqref{eq:Omgw0_sup}. The instrument noise $\OmInst(f_b)$ is described by Eq.~\eqref{eq:OmInst}.

Irrespective of the parameters on which the MCMC samples, the injected fiducial is calculated using the thermodynamic parameters as follows:
\ben
\Omega_{\textrm{fid}} =   \OmInst(f_b) + \Om_{\textrm{pt}}(f_b,\tilde{\th}_{\rm fid})
\een
are generated in the frequency domain using 1000 frequency logarithmic spaced points, $ \Om_{\textrm{pt}}(f_b,\tilde{\th}_{\rm fid})$  is described by Eq.~\eqref{eq:Omgw0_sup}  and $\OmInst$ by  Eq.~\eqref{eq:OmInst}. The injected power spectrum is a Gaussian draw around the theoretical fiducial model. In this work we do not consider any astrophysical foregrounds, as our focus is on the reconstruction of parameters. Furthermore, the fiducial models we go on to consider are strong enough that we expect the foregrounds to have little impact. For an exploration of the impact of foregrounds on LISA's ability to detect a SGWB from a first order phase transition see \cite{spec_separation_2022}.

\subsection{Priors on thermodynamic parameters}
\label{sec:thermopriors}
The priors on the the four thermodynamic parameters are chosen based on constraints from theory, simulations, the corresponding signal-to-noise ratio $\SNR$ of the GW signals they produce, and trustworthiness of the SSM.

The prior on the nucleation temperature $\Tn$ was chosen so the temperature scale is relevant to the electroweak scale. Due to the large range of scales involved, we impose a log-uniform prior between $\Tn = 10$ GeV -- $50$ TeV.

For the phase transition strength $\al$, which we remind the reader is the ratio of potential energy to thermal energy, we place a lower bound of $\al = 0.01$, which corresponds roughly to the lowest phase transition strength with signal-to-noise
ratio $\SNR>1$  for the $(r_*$, $\Tn)$ cases we consider. For the upper bound we use $\al = 0.67$, which is the highest phase transition strength used in current simulations \cite{Cutting:2019zws}. We impose a log-uniform prior for $\al$.

We place a log-uniform prior on the Hubble-scaled mean bubble spacing $r_*$ with a lower bound $r_* = 0.0005$, as lower signals are not observable i.e.\ $\SNR<1$ even for largest phase transition strength. The upper bound in general could be up to $r_*\simeq1$, otherwise the bubbles would be bigger than the observable universe. The SSM assumes the phase transition completes much faster than one Hubble time, which corresponds to $r_* \ll 1$. In practice we use an upper bound of $r_* = 0.5$.

Theoretically, the wall speed $\vw$ could take any value between 0 and 1 (where 1 indicates the speed of light in natural units).
Here we choose to use the current region explored by simulations and apply a flat uniform prior between $\vw = 0.24$  and $0.92$.

We also include a joint prior on $\al$ and $\vw$ that arises from the maximum phase transition strength  $\al_{\text{max}}$ for a given wall speed \cite{Espinosa:2010hh}. We use an approximate form of this relationship
\ben\label{eq:al_max}
\al_{\text{max}} = \frac{1}{3} \frac{\left(1 + 3 \vw^2\right)}{\left(1 - \vw^2\right)}.
\een
We summarise the priors on the thermodynamic parameters in Table~\ref{table:thermo_priors}.
\begin{table}[h!]
\centering
\begin{tabular}{|c|c|c|}
  \hline
 Parameter & Min & Max \\
  \hline
 $\log_{10}(\Tn/\mathrm{GeV})$ & $\log_{10}(10)$ &$\log_{10} (50 \times 10^3)$ \\
   \hline
 $\log_{10}\al$  & $\log_{10}(0.01) $ &$\log_{10} (0.67)$  \\
   \hline
 $\log_{10}r_*$  & $\log_{10}(0.0005)$  &$\log_{10}(0.5)$  \\
   \hline
 $\vw$  & $ 0.24$ &$0.92$  \\
\hline
\end{tabular}
\caption{ Ranges for the uniform priors on the thermodynamic parameters.}
\label{table:thermo_priors}
\end{table}

\subsection{Initial priors on spectral parameters }
\label{sec:spectralpriors}
The naive priors on the spectral parameters are chosen to allow for a generous spread around what we take to be observable, spectra with $\rho >1$. We do this to give the optimiser a wide range of spectral parameters when fitting to the thermodynamic parameters. The spectral priors are summarised in Table~\ref{table:spectral_priors}.
The prior on the break ratio $\rb$ is chosen to be linear as $\rb$ is closely related to the wall speed $\vw$, which has a linear prior.
For the intermediate slope $b$ we use a prior range that encompasses the range we found when fitting the double broken power law to a range of SSM spectra in \cite{Gowling:2021gcy}.
\begin{table}[h!]
\centering
\begin{tabular}{|c|c|c|}
  \hline
 Parameter & Min & Max\\
  \hline
 $\log_{10}\OmPeak$ & $\log_{10}(1\times 10^{-20}) $ &$ \log_{10}(1 \times 10^{-7})$  \\
 \hline
 $\log_{10}(\fp/\mathrm{Hz})$  & $\log_{10}(1 \times 10^{-7}) $ &$ \log_{10}(1)$\\
\hline
 $\rb$  & $1\times 10^{-7} $ &$1  $  \\
\hline
 $b$  & $-2 $ &$ 2$ \\
\hline
\end{tabular}
\caption{Ranges for the uniform priors on the spectral parameters.}
\label{table:spectral_priors}
\end{table}

\subsection{Markov chain Monte Carlo inference}

We sample from the likelihood described above, combined with different priors, using the adaptive Markov chain Monte-Carlo (MCMC) algorithm \cite{Lewis:2002ah} included in \texttt{Cobaya} \cite{Torrado:2020dgo}. The resulting chains are analysed using \texttt{GetDist} \cite{Lewis:2019xzd} in order to produce posterior density plots and credible intervals.

For each of our fiducial models we consider three set-ups: sampling on the spectral parameters $\theta=\spectralparamswprior$ with flat priors, sampling on the spectral parameters with the induced priors described in \ref{Sec:induced_prior} (in order to reconstruct the thermodynamic parameters), and finally, as a benchmark, sampling directly on the thermodynamic parameters $\tilde{\theta}=\thermoparamswprior$.

\section{Reconstructing thermodynamic parameter posteriors}
\label{Sec:recon}
To take advantage of the computationally cheaper double broken power law, we introduce a method for transforming spectral parameters into the corresponding thermodynamic parameters.  We generate a map $\Theta$ between the two parameter spaces by fitting the spectral parameters over a regular grid of thermodynamic parameters. This map is then used to generate an induced prior on the spectral parameters that is informed by our chosen thermodynamic parameter space. Finally, we introduce our reconstruction method using $\Theta$ and comment on the utility and interpretation of the reconstructed posterior.

\subsection{Constructing the map between spectral and thermodynamic parameters}
\label{Sec:min_fit}
Here we aim to make a map between spectral and thermodynamic parameters, as an analytic expression connecting the two sets of parameters does not exist.
We wish to find the spectral parameters giving the best fit for a GW power spectrum defined by a given set of thermodynamic parameters. 
To do that, we could use
least-squares curve fitting between the two spectra. Assuming that there will be imperfections in the mapping (e.g.\ regions in the thermodynamic parameter producing features that cannot be represented by the simpler spectroscopic template) there is a decision to be made about
which parts of the power spectra should be allowed to fit best. A natural prescription would be favouring the frequencies to which LISA is most sensitive, which could be implemented by weighting frequency bins during the fitting with the respective sensitivities.
We accomplish this with a
maximisation of the log-likelihood of Eq.~\eqref{eq:log_like}, where a thermodynamic template is injected as the fiducial model and a spectroscopic one is fitted to it. We use the optimiser code in \texttt{Cobaya} which uses \texttt{Py-BOBYQA} \cite{https://doi.org/10.48550/arxiv.1812.11343,https://doi.org/10.48550/arxiv.1804.00154}.
This defines the map and its numerical approximation.

We evaluate the map by using the above procedure to fit the gravitational wave power spectra for a regular 4D grid of thermodynamic parameters; each evaluation returns a vector of spectroscopic parameters.  These vectors are assembled into a 4D array of 4-component vectors, $\Theta$, which we refer to as the fit array.

The underlying regular grid of thermodynamic parameters is summarised in Table~\ref{table:minimzer_fit_array}.
The fact that it is regularly-spaced according to the uniform density of the set of thermodynamic parameters $\thermoparamswprior$ will make the computation of the induced prior simpler, as we will see below.

 The reader will note that the lower bounds for $\vw$, $r_*$, and $\Tn$  in the regular grid of thermodynamic parameters (see Table~\ref{table:minimzer_fit_array}) do not directly correspond to the ranges used for the priors in Table ~\ref{table:thermo_priors}.  As we will go on to consider high signal-to-noise ratio fiducial models we do not expect the MCMC chain to explore these relatively low signal-to-noise ratio regions. In order to reduce the computation time of the fit array and focus on a denser population of points in the regions of parameter space we expect the chains to explore, we trim the  lower bounds on $\vw$, $r_*$, and $\Tn$. 

\begin{table}[h!]
\centering
\begin{tabular}{|c|c|c|c|c|}
\hline
 Parameter & Min. & Max. & No.\ of points & Scale \\
  \hline
  $\Tn$ & $ 50 \rm GeV$ &$5000 \rm GeV$ &$ 20 $ &logarithmic   \\
  \hline
$\al$  & $0.01$ &$0.67$ &$ 44$ & logarithmic \\
 \hline
 $r_*$ & $ 0.05$ &$0.5$ &$ 19 $ &logarithmic   \\
  \hline
$\vw$  & $0.4 $ &$ 0.9$ &$ 43 $ &linear  \\
\hline
\end{tabular}
\caption{ Regular grid of thermodynamic parameters used to construct the fit array. Notice that the scaling corresponds to the prior density in Table \ref{table:thermo_priors}.}
\label{table:minimzer_fit_array}
\end{table}

The starting point for each fit was the following: $\OmPeak$: peak value for injected GW power spectra,  $\fp$: frequency corresponding to the peak amplitude, $\rb$: $0.5$, and $b$: $0.4$. The values for $\rb$ and $b$ were chosen to be generic starting points.  In order to improve efficiency of the fit array generation, we chose convergence criteria $\rho_{\text{end}}$,  which corresponds to the minimum allowed value of the trust region radius, to depend on signal-to-noise ratio $\rho$:
 \ben\label{eq:rho_end}
    \begin{array}{ccc}
         \rm \rho \leq0.001	\qquad &\rho_{\text{end}}&=0.01,\\
        0.001 < \rm \rho \leq 1  \qquad &\rho_{\text{end}}&=0.001,\\
       \rm \rho> 1 \qquad &\rho_{\text{end}}&=0.00001.\\
     \end{array}
\end{equation}

The fit array described here, which is a catalogue of thermodynamic parameters and their corresponding spectral parameters, forms the basis of both the theory-informed induced prior on the spectral parameters presented in Sec.\ \ref{Sec:induced_prior}, and our reconstruction algorithm presented in Sec.~\ref{Sec:thermorecon}.

The computational cost to generate the fit array can be split into two parts. Firstly,  we have to evaluate the theoretical GW power spectra for all parameter combinations, and then we have to perform the optimiser fits. For the SSM the first part is relatively quick because the GW power spectrum for different $r_*$ and $\Tn$ combinations can be rapidly evaluated by rescaling according to Eq.~\ref{eq:omgw_ssm}. The $718960$ optimiser fits to these spectra took $3000$ core hours and form the main upfront cost of the method. 

\subsection{Induced prior on the spectral parameters}
\label{Sec:induced_prior}
In this section we suppose that there is some physically motivated prior imposed on the thermodynamic parameters, $\pi(\tilde{\theta})$,
and address the problem of finding the prior induced by the map on the spectral parameters, $\pi(\theta)$. In the case that there are the same number of parameters $m$ in each space, and that the map is differentiable, the induced prior is the imposed prior multiplied by the Jacobian determinant of the map,
\ben
\pi(\theta) = \pi(\tilde{\theta}) \left| \frac{\partial(\tilde{\theta}_1,\ldots,\tilde{\theta}_m)}{\partial(\theta_1,\ldots,\theta_m)} \right|
\een
The Jacobian determinant gives the ratio between a volume element in the original $\tilde{\theta}$-space, and its image in the $\theta$ space.
Unfortunately, the mapping between spectroscopic and thermodynamic parameters is not analytic in most of the cases that we would consider in this context, so we must resort to the alternative approach. We have already obtained a sample of the thermodynamic parameters in the last section: the fit array $\Theta$. The density of the sample is proportional to the prior density, that is, both the grid and the prior are uniform in either the parameter or its logarithm.
Hence we can directly compute the prior for the spectral parameters as 
\begin{equation}
  \label{eq:spectralprior}
  \pi(\theta) = \Delta(\theta)\,,
\end{equation}
where $\Delta(\theta)$ is the density in the spectral parameter space induced by the mapping of the regular grid. As the fit array $\Theta$ is discrete, we use it to generate a frequency histogram on the spectral parameter space, and smooth the histogram value using a kernel density estimator (from \texttt{scipy.stats} \cite{2020SciPy-NMeth}) to approximate the density $\Delta(\theta)$. This is the prior that we will use in the MCMC runs which are aimed at recovering the thermodynamic from the spectral parameters. Notice that possible exclusion regions in the thermodynamic parameter space (such as that on the $(\al,\vw)$ described in Section \ref{sec:thermopriors}) are automatically accounted for in the mapping fit array $\Theta$, simply by the corresponding region having been excluded from the original grid.

\begin{figure}[h!]
    \centering
    \includegraphics[width=0.8\linewidth]{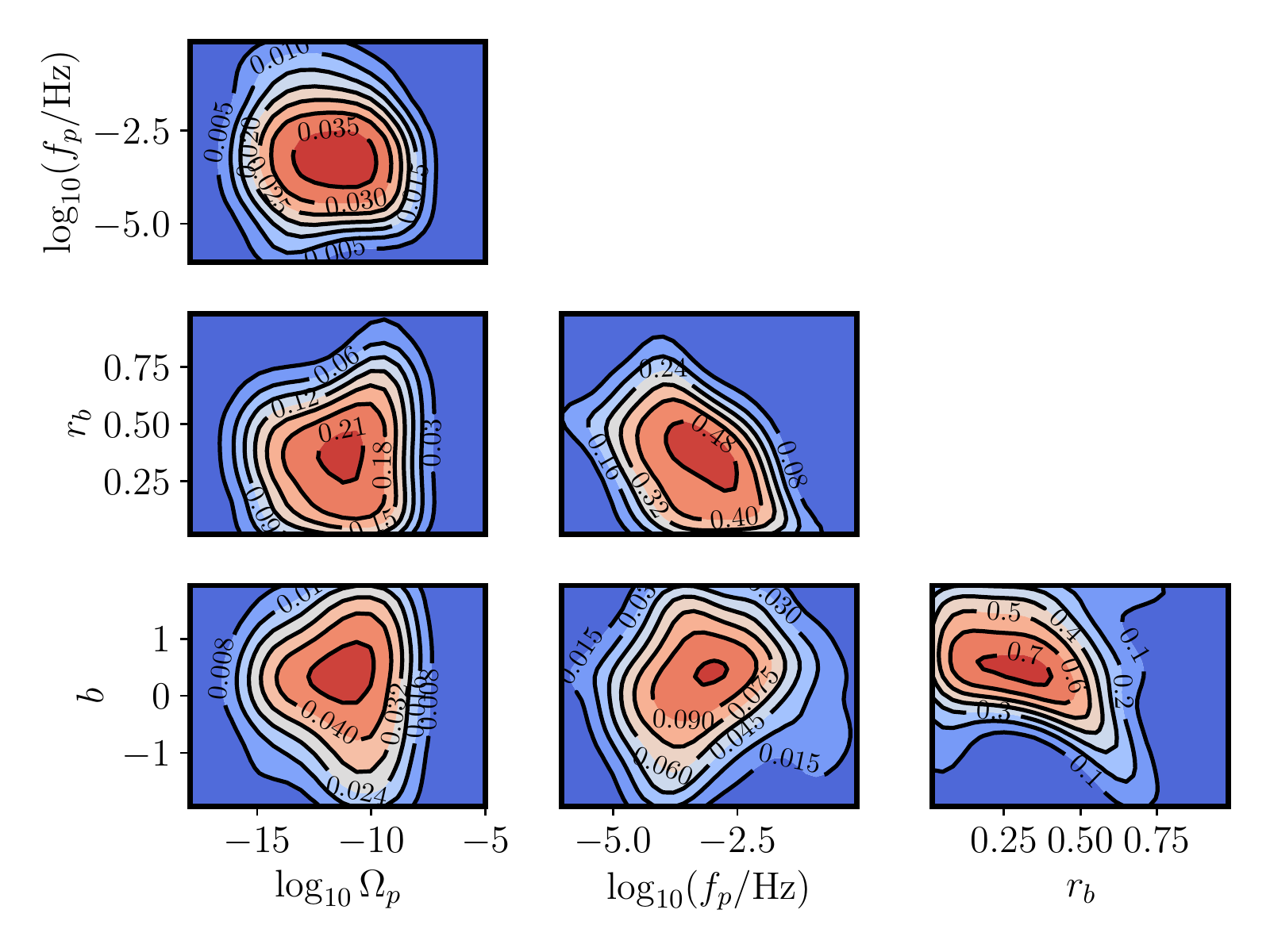}
    \caption{2D projections of the induced priors on the spectral parameters, where red and blue regions correspond to high and low probability respectively. Notice the difference between this prior density and the one described in Sec.\ \ref{sec:spectralpriors}, which remarks the need for the use of the induced prior in order for the recovered thermodynamic parameter constraints to be physically meaningful.}
  \label{Fig:induced_prior}
\end{figure}

The 2D projections of the induced prior probability density functions are shown in Fig.~\ref{Fig:induced_prior}, where the red and blue regions correspond to high and low prior probability respectively. The prior bounds we implement for the thermodynamic parameter space approximately correspond to the region of thermodynamic parameter space 
%observable by LISA (i.e.\ a 
where the SGWB has signal-to-noise ratio $\rho>1$. 
This means the induced priors shown in Fig.~\ref{Fig:induced_prior} contain the spectral parameter space for a first order phase transition observable at LISA. These priors are clearly different from the naive uniform priors that we started from for the spectroscopic parameters, i.e.\ uniform on $ \spectralparamswprior$. This difference remarks the need to account for the mapping by using the induced prior of Eq.\ \eqref{eq:spectralprior}, or we would be inadvertently imposing a very non-physical prior on the thermodynamic parameters when recovered as explained in the next section.

\subsection{Reconstruction of the thermodynamic parameters }
\label{Sec:thermorecon}
As a last step to produce constraints on the thermodynamic parameters from a sample of the spectral ones, we need to map the spectral parameters in the sample back to their corresponding thermodynamic ones. The fit array $\Theta$ cannot simply be inverted, since it is not regularly spaced in the spectral parameter space, and in any case we would need to interpolate to obtain mappings of arbitrary points that are not in the grid. Here we describe a procedure to do both the inversion and interpolation at once.

\begin{figure}[h!]
      \centering
       \includegraphics[width=0.8\linewidth]{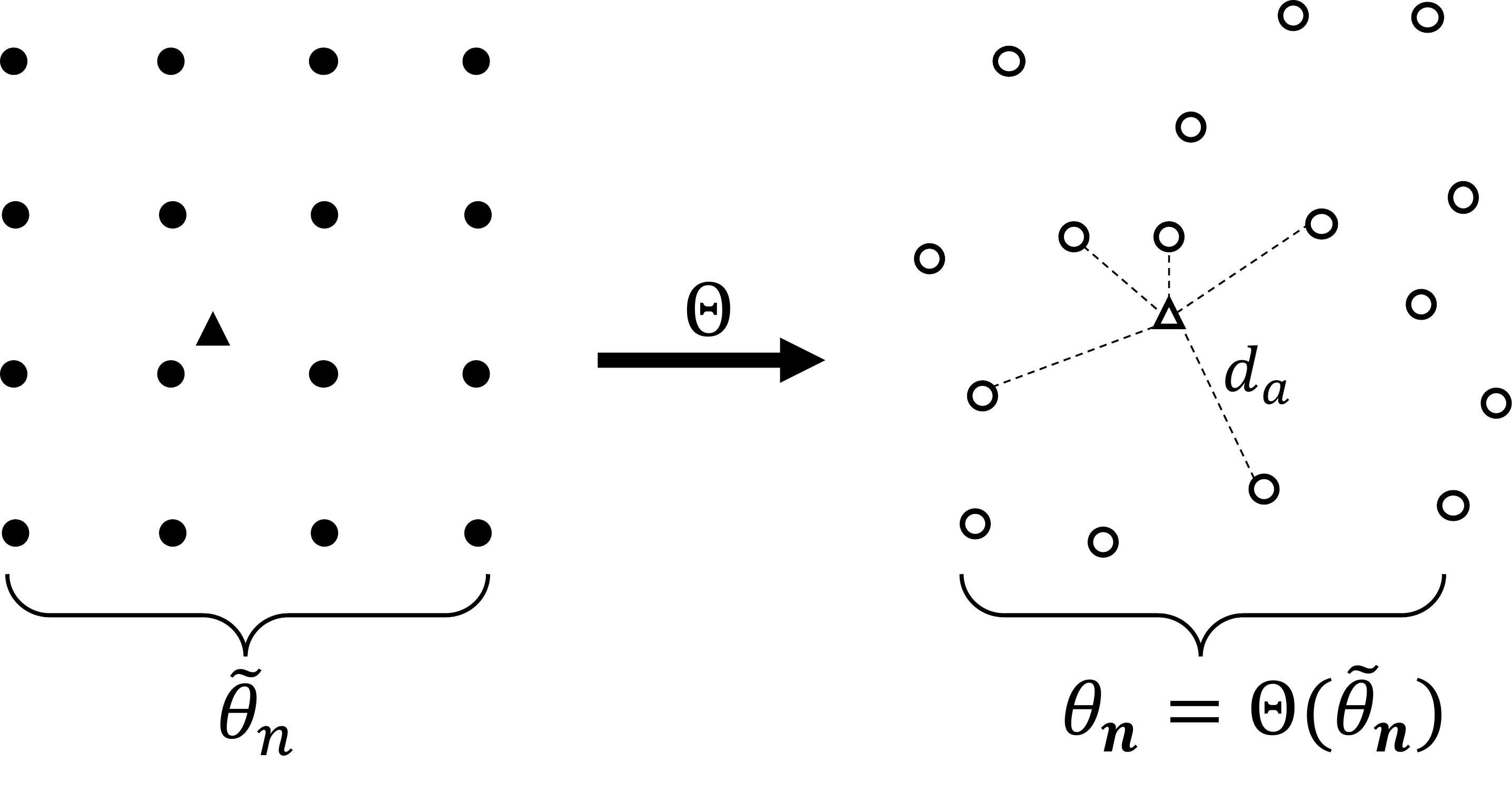}
\caption{{\label{Fig:recon_illustration}  A regular grid of thermodynamic parameters  $\tilde{\theta}_n$ shown
with filled points,  $\Theta$ is the fit array that connects the spectral parameters $\theta_n$ to the corresponding to  $\tilde{\theta}_n$. The irregular grid  of spectral parameters $\theta_n$ found using the optimiser fit are shown here as unfilled points.  $d_a$ is the distance between set of spectral parameters $\theta$ to reconstruct, shown here as a triangle, and one of the five nearest neighbours in the $\theta_n$ grid. The filled  triangle in the thermodynamic parameter space on the left represents the reconstructed thermodynamic parameters.}
}
\end{figure}

For a set of spectral parameters $\mathbf{\theta} = \spectralparamswprior$, the aim is to find a unique set of thermodynamic parameters $\tilde{\theta} = \thermoparamswprior$. 
We do this by finding a weighted nearest neighbour. 
The displacement of $\theta$ from a given element of $\Theta$ is given by
\ben\label{eq:delta_theta}
\Delta \theta = \theta - \Theta = (\Delta\log_{10}\OmPeak , \Delta \log_{10}\fp, \Delta \rb, \Delta b).
\een
%where $\Theta$ is the fit array connecting spectral and thermodynamic parameters generated using the optimiser fit function. 
The distance $d$, in spectral parameter space, between the input point $\theta$ and a point in the fit array $\Theta$ is given by
\ben\label{eq:rel_d_spectral}
d=  \sqrt{\Delta \log_{10} \OmPeak ^2  + \Delta\log_{10} \fp^2 + \Delta \rb ^2 + \Delta b^ 2} + \ep,
\een
where $\ep$ is a small value used as a regulator preventing divide by zero errors.  
We take the 5 smallest values of distances
to build the array $d_a$ of the 5 nearest neighbours.  The  5 corresponding sets of thermodynamics parameters $\tilde{\theta}_a$ are then
averaged with the inverse square of the distance (Eq.~\eqref{eq:rel_d_spectral})
\ben\label{eq:weighted_nearest}
\tilde{\theta} = \frac{ \sum_{a=1}^{N} {\tilde{\theta}_a}/{d_a ^2}}{\sum_{a=1}^{N}{1}/{d_a ^2}}.
\een
This is the reconstructed thermodynamic parameter, illustrated by the filled triangle in Fig.~\ref{Fig:recon_illustration}. The evaluation time of the reconstructed parameters as described in Eq.~\ref{eq:weighted_nearest} method is minimal so we can calculate them as we sample on the spectral parameters.

An important feature of this reconstruction technique is that excluded regions of the parameter space can never be accessed by the reconstructed parameters (as long as the mapping is well-behaved, which it will be if the spectroscopic template is a good enough approximation to the physical one). As the grid will not contain points in the excluded regions, all points $\tilde{\theta}_a$ corresponding to the nearest neighbours will necessarily be allowed values, and (provided the allowed region is convex) their weighted sum will too.

\subsection{Properties of the reconstructed thermodynamic posterior}
It would be desirable if the induced prior on the spectral parameters would approach the original thermodynamic prior when reconstructing the thermodynamic parameters on finer and finer grids. However, in order for this to be achievable, every possible physical template must be reproduced exactly by the spectroscopic template with some unique combination of the spectral parameters, the mapping between the two sets of parameters must be one-to-one,
and the optimiser must find the precise correspondence every time.

These conditions are not generally satisfied, and so it is to be expected that 
the recovered thermodynamic parameters will not be distributed according to the exact physical prior, and thus the reconstructed posteriors will not be equivalent to the ones we would obtain by sampling directly on the thermodynamic parameters. Nevertheless, reasonably small deviations from these conditions  (e.g.\ the spectroscopic template may miss some corner-case physical features, the fit array grid is fine but finite, or the optimiser fails to
find the best fitting function) will still produce priors with useful properties: parameter values for physically excluded regions can never be recovered (as explained in the last section), the base density for the thermodynamic parameters (e.g.\ uniform, log-uniform...) is preserved; and on data containing an actual signal, the best-fit model of a hypothetical thermodynamic sample has high likelihood of being contained within the reconstructed contours.

The inevitable differences in the prior indicates that the reconstructed posteriors should not be interpreted as a direct reconstruction of the actual ones, but these nice properties guarantee that they provide a sound but much cheaper first order approximation to parameter constraints in the physical parameters, which is physically reasonable (reproduces exclusions and densities) that can be used e.g.\ to refine the spectroscopic formula or the fit array in the region of interest to get an even better approximation.

In the next section we will find some of these differences and test the soundness of the reconstructed posteriors using to benchmark cases.
\section{Results}
\label{sec:results}

We perform MCMC inference for two fiducial models: a deflagration and a detonation, each with signal-to-noise  $\rho \sim 40$. In each case the injected signal contains the SGWB from a first order phase transition, as described by the SSM Eq.~\eqref{eq:Omgw0_sup}, and the LISA instrument noise, as described by Eq.~\eqref{eq:OmInst}.

For each fiducial model we perform three MCMC runs: sampling on the spectral parameters with the flat priors given in Table~\ref{table:spectral_priors}, sampling on the spectral parameters with the induced priors as described in Sec.~\ref{Sec:induced_prior} and Fig.~\ref{Fig:induced_prior}, and sampling on the thermodynamic parameters with the priors given in Table~\ref{table:thermo_priors}.

The  MCMC runs  are implemented using \texttt{Cobaya} \cite{Torrado:2020dgo} with a log-likelihood described by Eq.~\eqref{eq:log_like}. For the MCMCs that sample on the spectral parameters the SGWB from a phase transition is described by  Eq.~\eqref{eq:omgw_dbl_brkn} and $\theta = \spectralparamswprior$. When sampling directly on the thermodynamic parameters  the phase transition signature is described by Eq.~\eqref{eq:Omgw0_sup} and $\tilde{\theta}=\thermoparamswprior$.

The set-up for the MCMC runs is as follows.  \texttt{Cobaya} uses the Gelman-Rubin statistic $R -1$ as the convergence criteria, specifically we use $R -1\leq0.001$ for the spectral samples and $R-1\leq0.01$ for the thermodynamic samples (as they take longer to evaluate). The maximum number of tries at each point in the chain is $100000$. For the runs on spectral parameters we use the optimiser fit (as described in Sec.~\ref{Sec:min_fit}) to the injected phase transition signal as the starting point of the chain.
For the thermodynamic samples the starting point of the chain is taken from a Gaussian draw centred around the fiducial model values.

For the MCMC samples  performed here with a single chain and four threads, the spectral sample with the induced prior took $\sim$5 days to converge with $\sim$200,000 points in the chain. The corresponding direct sample on the thermodynamic parameters took $\sim$16 days to reach $R-1\leq0.01$ with $\sim$70,000 points in the chain.

\subsection{Deflagration fiducial model }
For the deflagration fiducial model we use $\vw = 0.55 $ , $\al = 0.4$, $r_* = 0.1$ and $\Tn = 120 $ GeV, which has a signal-to-noise ratio $\rho = 40.1$.

\begin{figure}[th!]
    \centering
  \begin{subfigure}{.49\textwidth}
    \centering \includegraphics[width=\linewidth]{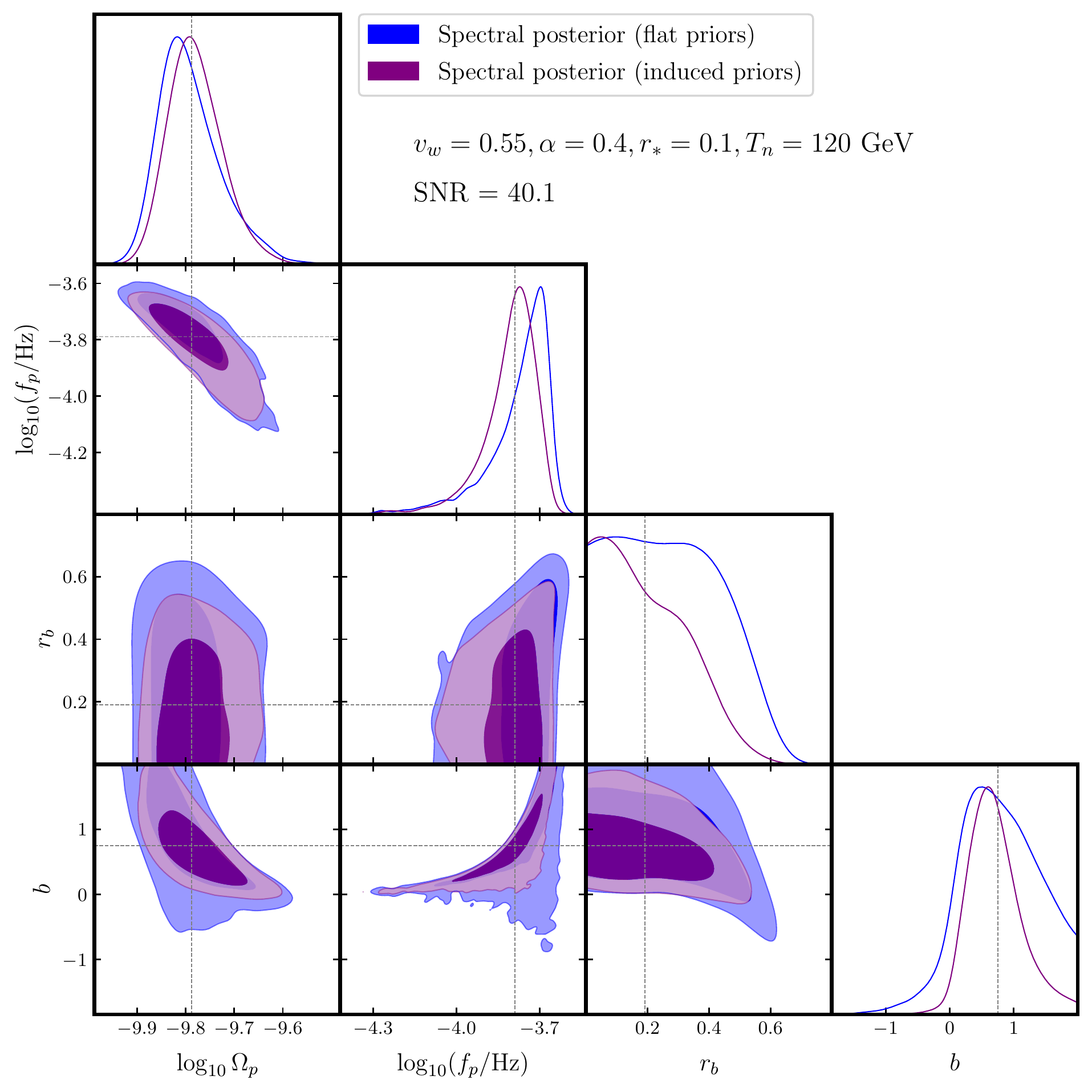}
    \caption{\label{Fig: spectral_def_triangle} }
  \end{subfigure}
  \begin{subfigure}{.49\textwidth}
    \centering \includegraphics[width=\linewidth]{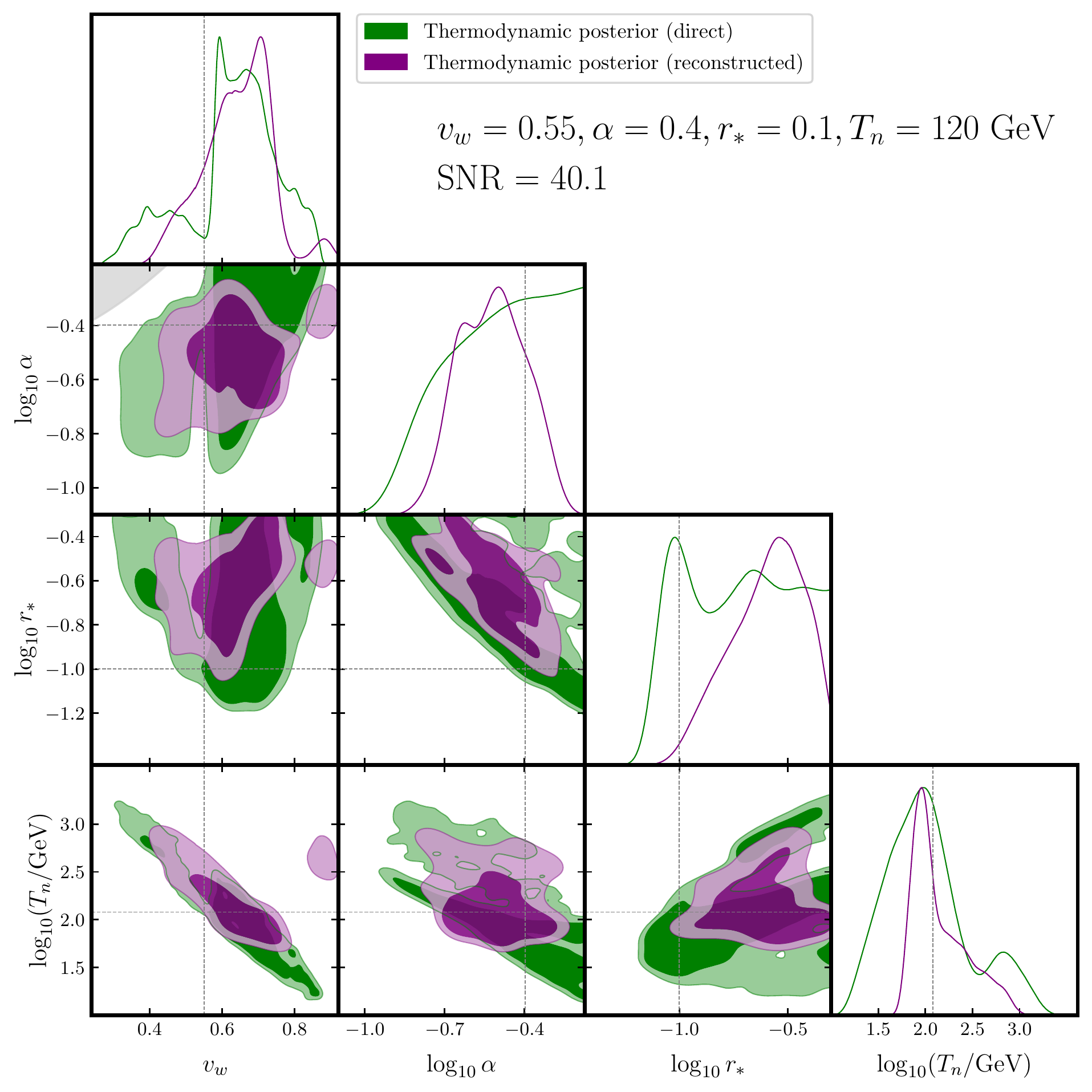}
    \caption{\label{Fig: thermo_def_triangle}  }
  \end{subfigure}

\caption{{\label{Fig:SNR_40_deflagration} Triangle plots for the deflagration fiducial model $\al = 0.4$, $\vw = 0.55$, $r_* = 0.1$, $\Tn =120$ GeV, for MCMCs sampling on spectral parameters \ref{Fig: spectral_def_triangle} and thermodynamic parameters \ref{Fig: thermo_def_triangle}.  On the left, the spectral MCMC samples with flat priors (blue) and with induced priors (purple). The cross hairs in the spectral triangle plots mark the best fit to the injected spectrum calculated using the optimisation procedure described in Section \ref{Sec:recon}.  On the right are the corresponding samples on the thermodynamic parameters (green) and thermodynamic parameters reconstructed from the spectral sample  (purple). The cross hairs in the thermodynamic triangle plot show the injected thermodynamic parameters.   The grey shading in the $\vw$-$\al$ plot shows the region excluded by the physical prior, described in Eq.~\eqref{eq:al_max}.
}}
\end{figure}

In Fig.~\ref{Fig: spectral_def_triangle} we present the results for the spectral samples in the deflagration case. The blue regions show the posterior with the uniform spectral priors given in Table~\ref{table:spectral_priors}. The purple regions show the posteriors when the induced prior informed by the thermodynamic parameter space is included. The cross-hairs show the start point of the chain, which corresponds to the optimiser fit to the spectrum generated from the thermodynamic fiducial model.

For each point in the spectral parameter chain with the induced priors we perform the reconstruction algorithm Eq.~\eqref{eq:weighted_nearest} to build a corresponding chain of reconstructed thermodynamic parameters. The distributions corresponding to reconstructed thermodynamic parameters are shown in purple in Fig.~\ref{Fig: thermo_def_triangle}.  The posterior from the MCMC sampling  directly on the thermodynamic parameters is shown in green in Fig.~\ref{Fig: thermo_def_triangle}.

The  marginalised 1D and 2D posteriors in Fig.~\ref{Fig: spectral_def_triangle} from the flat and induced priors are in good agreement. The 2D posteriors for the MCMC sample with the induced priors (purple) cover a smaller area. The difference is mostly due to a prior cut for large $\rb$, which is not favoured by the SSM, despite being allowed by the data.  The disfavouring of large $\rb$ values can be seen as a sharp fall for $\rb \simeq 0.5$ in the induced prior of Fig.\ \ref{Fig:induced_prior}. The break ratio $\rb$ is the hardest for the MCMC to estimate as it requires knowledge of both breaks in the GW power spectrum. In this case (and in general) one of the breaks is at low or high frequencies and out of LISA's peak sensitivity region. The means and 68\% credible intervals for the spectral parameters for the flat and induced priors are summarised in Table~\ref{table:spectral_deflagration_bounds}.

\renewcommand{\arraystretch}{1.2}
\begin{table}[h!]
\centering
\begin{tabular}{|c|c|c|c|c|}
\hline
  & $\log_{10} \OmPeak $ &$ \log_{10} (\fp /\rm Hz)$ &$ \rb$ &$ b $\\
  \hline
Flat priors  & $-9.791^{+0.044}_{-0.075} $ &$ -3.78^{+0.12}_{-0.036}$ &$  < 0.368 $ &$ 0.78^{+0.58}_{-0.67} $ \\
 \hline
 Induced priors & $ -9.779^{+0.046}_{-0.063}$ &$-3.81^{+0.11}_{-0.048}$ &$  < 0.267 $ &$ 0.70^{+0.30}_{-0.47}$   \\
\hline
\end{tabular}
\caption{ Means and 68\% credible intervals for the spectral parameters, deflagration fiducial model. }
\label{table:spectral_deflagration_bounds}
\end{table}
\renewcommand{\arraystretch}{1.0}

We now consider the results for the posteriors on the thermodynamic parameters and compare
the results from the direct sample and the reconstructed sample. In Fig.\ \ref{Fig: thermo_def_triangle} there is general agreement between the two sets of 2D posteriors. 
 In particular, we note the directions of the correlations in the 2D posteriors are recovered well in the reconstructed sample. The largest difference appears for the Hubble-scaled mean bubble spacing $r_*$, which has a tighter lower bound and more defined peak than the posterior from the directly sampled thermodynamic parameters. This difference is not surprising, since the direct thermodynamic sample also fails to recover $r_*$.  This is because the injected $r_*$ value is hard to distinguish from higher ones:  the SGWB for this deflagration has a plateau peaking at a frequency lower than LISA's peak sensitivity, and increasing $r_*$ displaces the signal peak towards lower frequencies at the same time as increasing the amplitude, keeping the signal-to-noise approximately constant (see Fig.\ 1c of \cite{Gowling:2021gcy}). This effect can also be seen as a degeneracy between $\OmPeak$ and $\fp$; 
the reconstruction simply selects from the long tails the values that are more likely to be reproduced by a spectroscopic template. The wall speed posterior is bi-modal  because away from the speed of sound detonations and deflagrations have similar spectral shape (this can be seen in Fig.~1 in \cite{Gowling:2021gcy}). 

The means and 68\% credible intervals for the thermodynamic parameters for the direct and reconstructed samples  for the deflagration case are summarised in Table~\ref{table:thermo_deflagration_bounds}.

\renewcommand{\arraystretch}{1.2}
\begin{table}[h!]
\centering
\begin{tabular}{|c|c|c|c|c|}
  \hline
  & $ \vw$ &$ \log_{10} \al$ &$ \log_{10} r_*$ &$\log_{10}  (\Tn/\rm GeV) $\\
  \hline
Fiducial model    & $0.55$ &$-0.398$ &$-1 $ &$2.08$ \\
  \hline
Direct  & $0.630^{+0.17}_{-0.059}$ &$ > -0.595$ &$> -0.890 $ &$2.03^{+0.27}_{-0.54}$ \\
 \hline
 Reconstructed & $0.646^{+0.098}_{-0.075}$ &$-0.52^{+0.12}_{-0.15}$ &$-0.59^{+0.22}_{-0.13} $ &$ 2.15^{+0.14}_{-0.36} $   \\
\hline
\end{tabular}
\caption{
Thermodynamic parameters for the fiducial deflagration model, and the thermodynamic parameters inferred from the MCMC samples.
``Direct'' uses chains sampled directly on the thermodynamic parameters, ``reconstructed''  uses chains sampled on the spectral parameters,
and reconstructs the corresponding thermodynamic parameters using the method described in Section \ref{Sec:recon}.
Values given are means and 68\% confidence intervals.
}
\label{table:thermo_deflagration_bounds}
\end{table}
\renewcommand{\arraystretch}{1.0}

\begin{figure}[h!]
    \centering
    \begin{subfigure}{.45\textwidth}
    \centering
    \includegraphics[width=\linewidth]{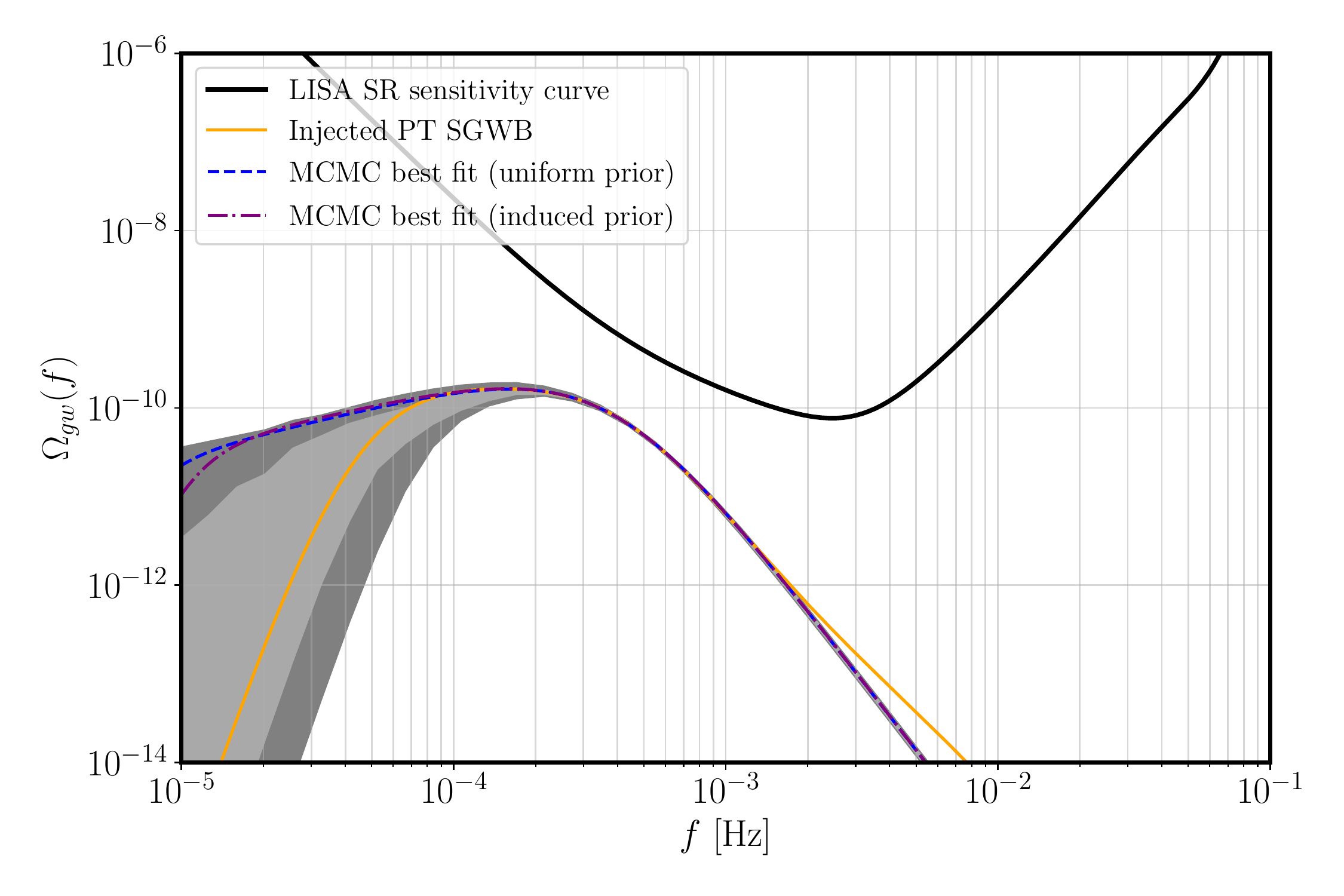}
    \caption{\label{Fig: spectral_def_spectra}  }
  \end{subfigure}
\hfill
  \begin{subfigure}{.45\textwidth}
    \centering
    \includegraphics[width=\linewidth]{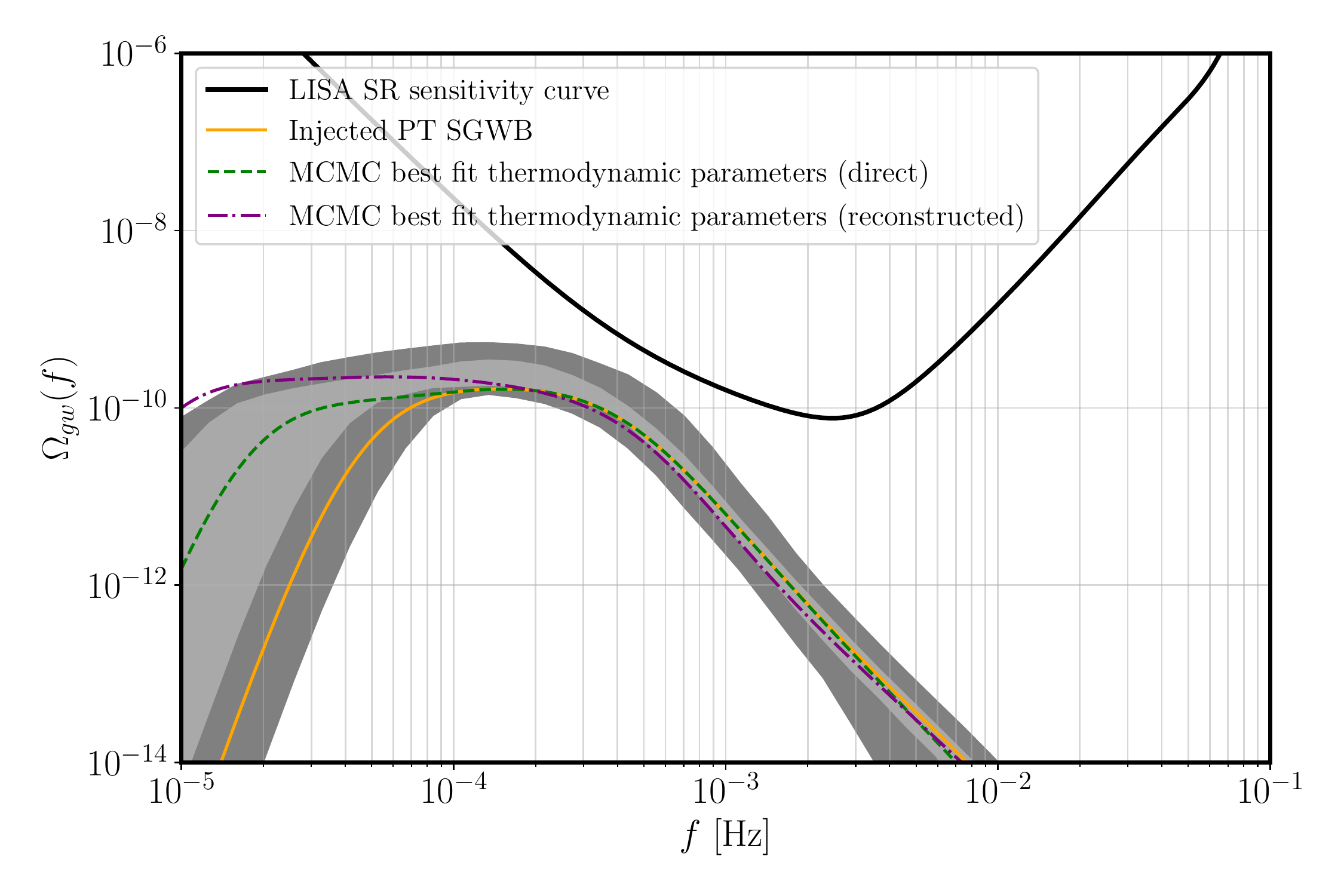}
    \caption{\label{Fig: thermo_def_spectra}  }
  \end{subfigure}
  \caption{\label{Fig:SNR_40_deflagration_Spectra} 
  Injected and best fit spectra for the detonation fiducial model with $\vw = 0.88$, $\al = 0.2$, $r_* = 0.1$, $\Tn = 200$GeV. The light and dark grey bands show the $1$ and $2$ sigma spread on the power spectra for the MCMC sample with the induced prior. In the spectral parametrisation (a) the best fit spectrum with the uniform prior is shown in blue, and the induced prior is shown in purple. In the thermodynamic parametrisation (b): the best fit spectrum for the direct sampling is shown in green, and the reconstructed sampling in purple. In both cases the injected spectrum is shown in yellow.
}
\end{figure}

In Fig.~\ref{Fig: spectral_def_spectra} we compare GW power spectra for the injected deflagration fiducial model (orange line) with the best fit spectra for the MCMC inferences, with flat and induced priors on the spectral parameters, shown in blue and purple respectively. The light grey and dark grey bands highlight the $68\%$ and $95\%$ confidence intervals on the GW spectra from the MCMC simulation which samples on the spectral parameters with the induced prior. In the frequency window that corresponds to LISA's peak sensitivity the spectra agree well. In the low frequency region the best fit for the induced prior run does not match with the injected phase transition signal so well; here LISA has little constraining power because of the low sensitivity, and the induced prior 
does not prevent sampling on very low values of $\rb$.  For the MCMC on the thermodynamic parameters the best fit spectra are shown in purple and green for the reconstructed and direct samples respectively  in Fig.~\ref{Fig: thermo_def_spectra}. Here we see the spectrum from the best fit of the reconstructed thermodynamic parameters sample falls within the  95\% confidence band over the majority of the frequency band.

\subsection{Detonation fiducial model}
For the detonation fiducial model we use $\vw = 0.88 $ , $\al = 0.3$, $r_* = 0.1$ and $\Tn = 200 $ GeV, which has a signal-to-noise ratio $\rho = 38.6$. In this case, the chosen wall speed is close to the upper bound on the prior, and so we expect this choice to test the edge effects in the reconstruction method.

\begin{figure}[h!]
        \centering
    \begin{subfigure}{.49\textwidth}
    \centering    \includegraphics[width=\linewidth]{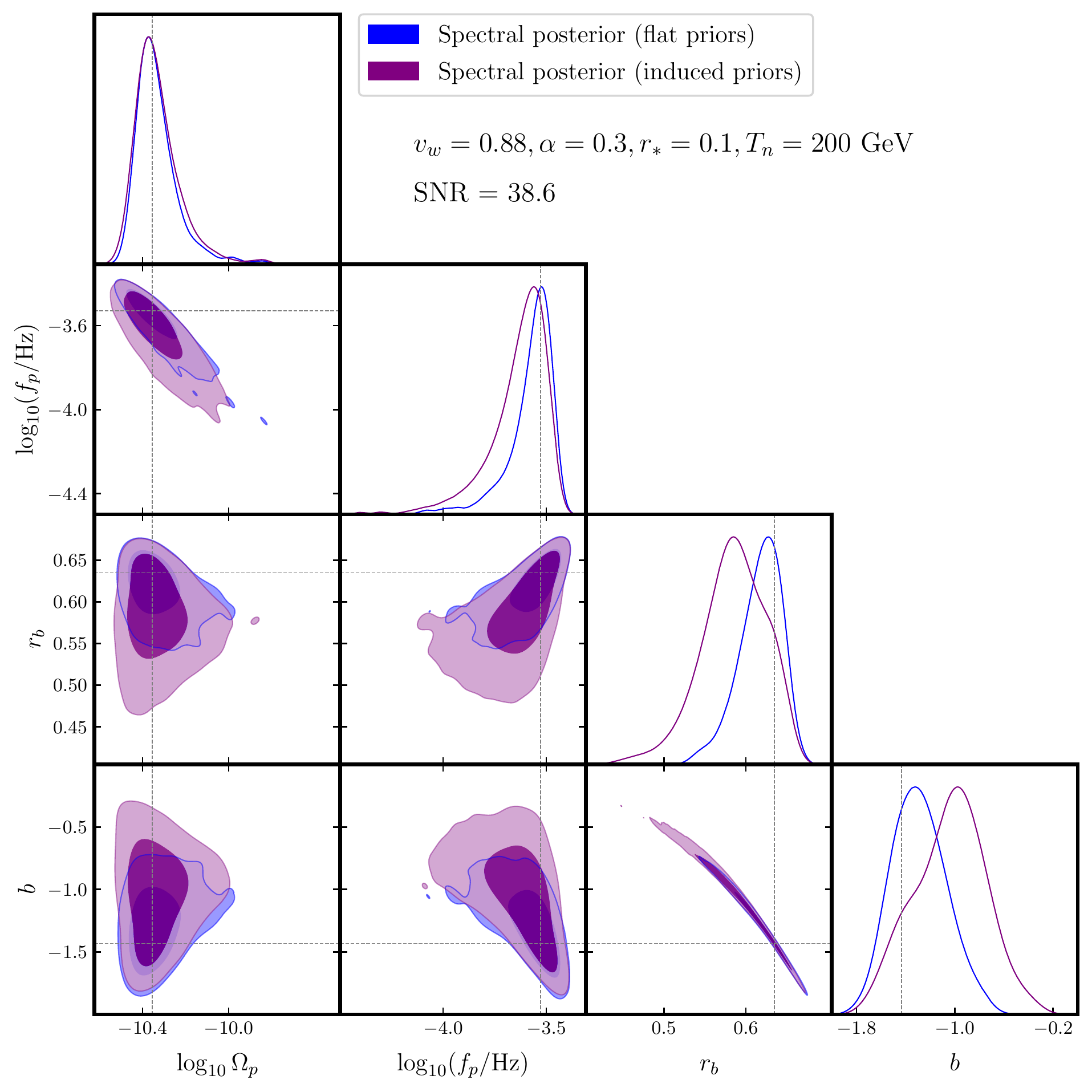}
    \caption{\label{Fig:spectral_det_triangle} }
  \end{subfigure}
  \begin{subfigure}{.49\textwidth}
    \centering \includegraphics[width=\linewidth]{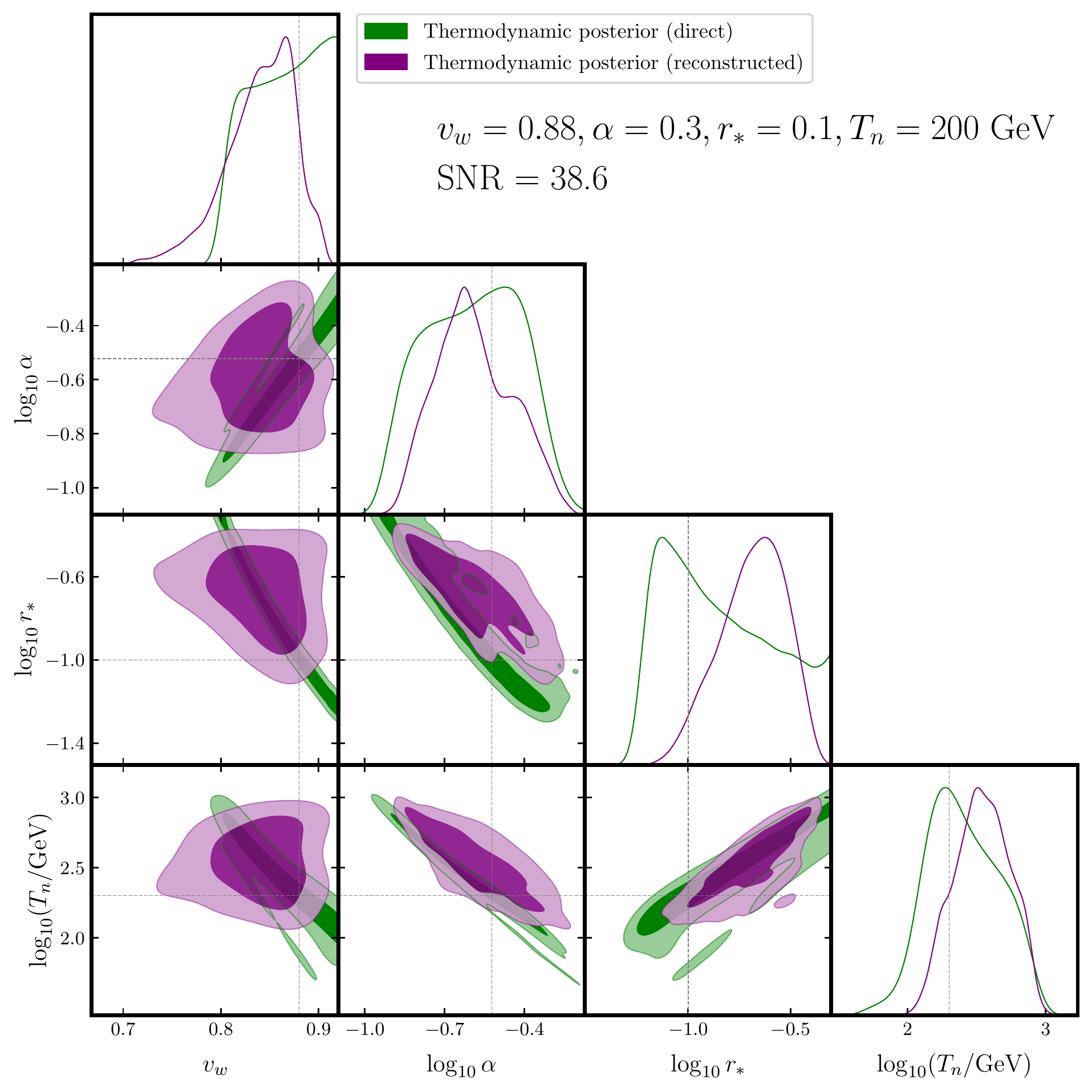}
    \caption{\label{Fig: thermo_det_triangle} }
  \end{subfigure}
   \caption{{\label{Fig:SNR_40_detonation_runs} Same as Fig.~\ref{Fig:SNR_40_deflagration} but for the detonation fiducial model with $\vw = 0.88$, $\al = 0.2$, $r_* = 0.1$, $\Tn = 200$ GeV. }
}
\end{figure}

We follow the same approach for the detonation as for the deflagration fiducial model. In Fig.~\ref{Fig:spectral_det_triangle} we present the triangle plots for the spectral parameters with flat priors (blue) and induced priors (purple). Again, there is good agreement between the 1D and 2D posteriors from the flat and induced priors. Here, unlike the deflagration case, the 2D posteriors for MCMC runs with the induced priors cover a larger area than those for the flat priors. In this case the spectral best fit has a large negative $b$, which is disfavoured by the induced prior, so the sampling is predominantly on less negative values of $b$.  The strong correlation between $b$ and $\rb$ increases the apparent area wherever one of these parameters appears. The means and 68\% credible intervals of the chains are presented in Table~\ref{table:spectral_detonation_bounds}.

\renewcommand{\arraystretch}{1.2}
\begin{table}[h!]
\centering
\begin{tabular}{|c|c|c|c|c|}
\hline
  & $\log_{10} \OmPeak $ &$ \log_{10} (\fp /\rm Hz)$ &$ \rb$ &$ b $\\
  \hline
Flat priors  & $-10.332^{+0.050}_{-0.11} $ &$-3.58^{+0.12}_{-0.042}$ &$0.617^{+0.031}_{-0.019}$ &$-1.29^{+0.20}_{-0.25}  $ \\
 \hline
 Induced priors & $-10.326^{+0.057}_{-0.12} $ &$ -3.64^{+0.16}_{-0.059} $ &$0.585^{+0.047}_{-0.033}$ &$-1.04^{+0.33}_{-0.28}$   \\
\hline
\end{tabular}
\caption{Means and 68\% credible intervals for the spectral parameters, detonation fiducial model.
}
\label{table:spectral_detonation_bounds}
\end{table}

\renewcommand{\arraystretch}{1.0}

\begin{figure}[th!]
    \hfill
  \begin{subfigure}{.49\textwidth}
    \centering
    \includegraphics[width=\linewidth]{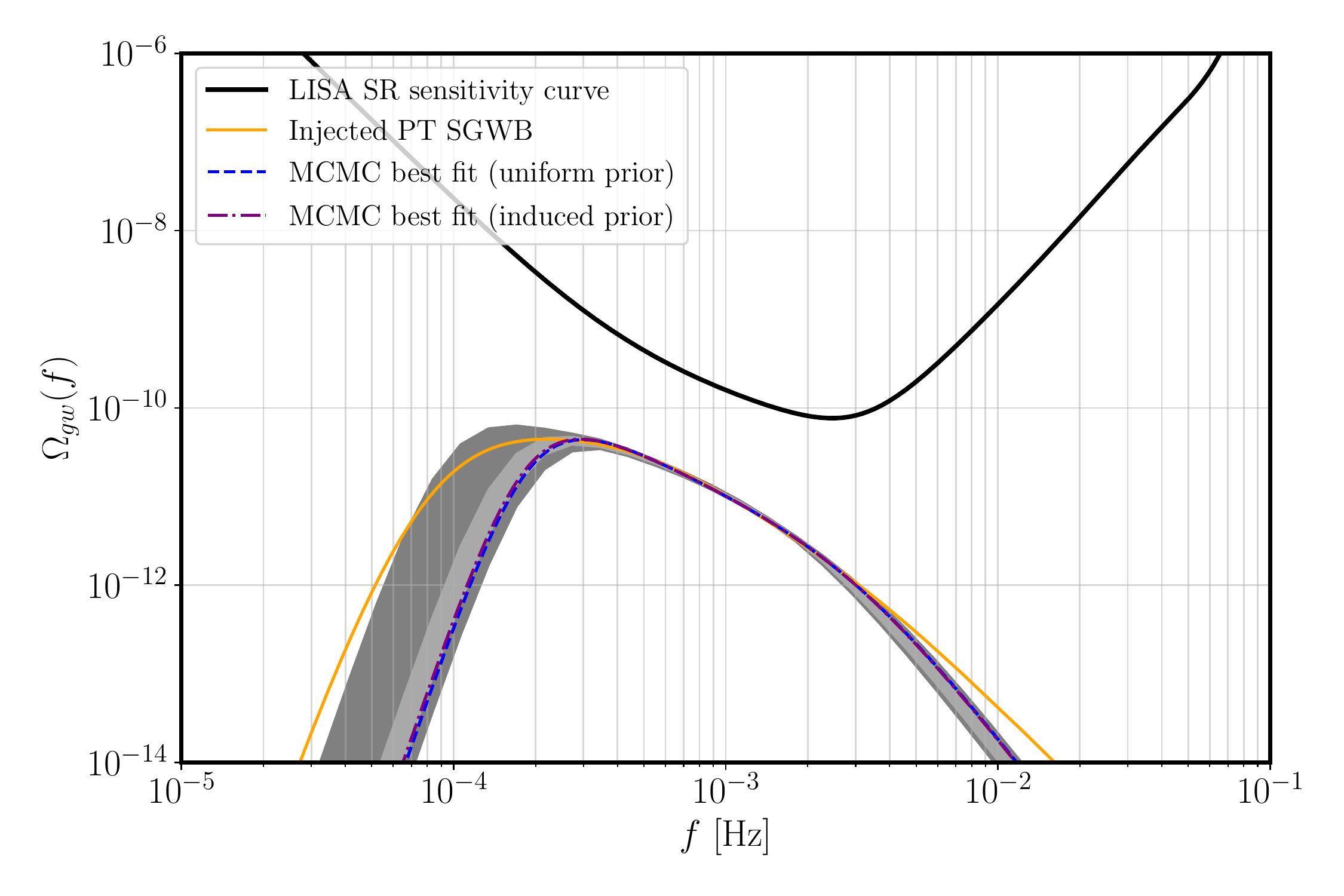}
    \caption{\label{Fig:spectral_det_spectra} }
      \end{subfigure}
  \hfill
  \begin{subfigure}{.49\textwidth}
    \centering
    \includegraphics[width=\linewidth]{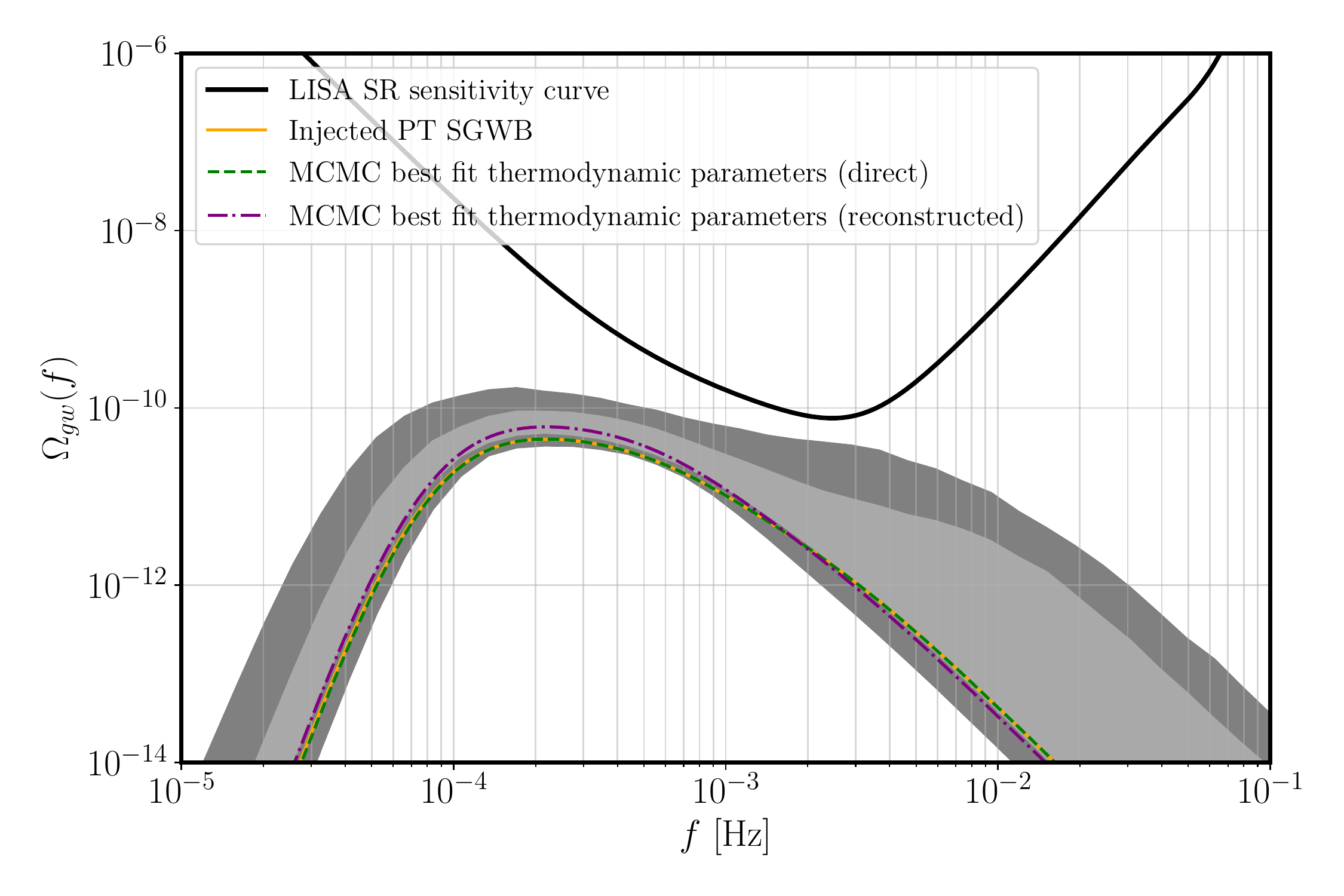}
    \caption{\label{Fig: thermo_det_spectra} }
  \end{subfigure}
\caption{\label{Fig:SNR_40_detonation_runs_spectra} Injected and best fit spectra for the detonation fiducial model with $\vw = 0.88$, $\al = 0.2$, $r_* = 0.1$, $\Tn = 200$GeV. 
The light and dark grey bands show the $1$ and $2$ sigma spread on the power spectra for the MCMC sample with the induced prior. In the spectral parametrisation (a) the best fit spectrum with the uniform prior is shown in blue, and the induced prior is shown in purple. In the thermodynamic parametrisation (b): the best fit spectrum for the direct sampling is shown in green, and the reconstructed sampling in purple. In both cases the injected spectrum is shown in yellow.
}
\end{figure}

In Fig.~\ref{Fig: thermo_det_triangle} we present the triangle plot for the reconstructed (purple) and direct (green) samples on the thermodynamic parameters. Again there is general agreement  in the 1D and 2D posteriors. 
For the 2D posteriors that include the wall speed there is a greater spread in the reconstructed samples, as the correlations with the other three parameters are not reproduced. This is a rather special region of thermodynamic parameter space: small changes in the wall speed make large changes in the power spectrum at higher frequencies, as can be seen in Fig.~1a of Ref.~\cite{Gowling:2021gcy}.  This accounts for the broadness of the 68\% and 95\% bands of the gravitational wave power spectrum for the reconstructed sample in Fig.~\ref{Fig: thermo_det_spectra}.
The GW power spectra possible in this region of parameter space are also not well described by the double broken power law. This results in a wide range of wall speeds being mapped onto the small range of spectral parameters in the fit array, which subsequently causes the broad spread in the reconstructed posteriors.

An edge effect is also on display in the 1D and 2D posteriors for $\vw$: the sampling on the thermodynamic parameters explores the region all the way up to the upper bound, while there is a cut-off in the posterior reconstructed from the sampling with spectral parameters.  This can be ascribed to the kernel density estimate smoothing the prior at the boundaries. We would expect to reduce the edge effect by refining the grid near the boundary.

The means and 68\% credible intervals for the thermodynamic parameters in the case of  the detonation fiducial model are displayed in Table~\ref{table:thermo_detonation_bounds}.

\begin{table}[h!]
\centering
\begin{tabular}{|c|c|c|c|c|}
  \hline
  & $ \vw$ &$ \log_{10} \al$ &$ \log_{10} r_*$ &$\log_{10} (\Tn/\rm GeV) $\\
    \hline
  Fiducial model   & $0.88$ &$-0.52$ &$-1$ &$2.30 $ \\
  \hline
  Direct  & $> 0.843$ &$-0.60^{+0.21}_{-0.17}$ &$-0.85^{+0.18}_{-0.37}$ &$2.41^{+0.25}_{-0.31}  $ \\
 \hline
 Reconstructed &$0.840^{+0.041}_{-0.025} $ &$ -0.59^{+0.15}_{-0.18} $ &$-0.68^{+0.20}_{-0.13}  $ &$2.54^{+0.23}_{-0.20} $   \\
\hline
\end{tabular}
\caption{ Thermodynamic parameters for the fiducial detonation model, and the thermodynamic parameters inferred from the MCMC samples.
``Direct'' uses chains sampled directly on the thermodynamic parameters, ``reconstructed''  uses chains sampled on the spectral parameters,
and reconstructs the corresponding thermodynamic parameters using the method described in Section \ref{Sec:recon}.
Values given are means and 68\% credible intervals.
\label{table:thermo_detonation_bounds}
}
\end{table}

In Fig.~\ref{Fig:spectral_det_spectra} we show the $68\%$ and $95\%$ confidence band on the GW spectra from the MCMC simulation which samples on the spectral parameters with the induced prior. The injected signal falls within the $95\%$ confidence band in both the spectral parametrisation and in the thermodynamic parametrisation.  In the thermodynamic parametrisation, shown in Fig.~\ref{Fig: thermo_det_spectra}, the best fit spectra 
 coincide very well with the injected spectrum, for both direct (green) and reconstructed sampling (purple).

\FloatBarrier
\section{Conclusions}
\label{sec:conclusions}

In this paper we introduced and tested a method for investigating LISA's sensitivity to a SGWB from a first order phase transition using parametrised templates as an approximation to the 
more complete sound shell model (SSM) of gravitational wave production. 
The parametrised template took the form of a double broken power law, a function of four ``spectral'' parameters. 
We investigated what information about the thermodynamic parameters of the sound shell model (wall speed $\vw$, phase transition strength $\al$, Hubble-scaled mean bubble spacing $r_*$, and nucleation temperature $\Tn$) can be obtained from sampling on the spectral parameters. 
The double broken power law is advantageous as it is a simple function, and therefore much faster to evaluate than the SSM, which involves a rather complex sequence of operations \cite{Hindmarsh:2019phv}. 
However, the mapping from the spectral to the thermodynamic parameters is not straightforward, as discussed in  \cite{Gowling:2021gcy}. Here we have proposed a reconstruction method as a solution to this problem. 

The motivation for developing the reconstruction algorithm was to provide a less computationally intense way to perform MCMC runs that constrain the thermodynamic parameters of a first order phase transition. The evaluation time of a proposal in the reconstructed chain is $\mathcal{O}(1000)$ times quicker than in the direct chain.  This reconstruction method could be applied to other data analysis problems where the connection between a computationally intensive theoretical model and an analytic fit is required.

A key component of the  reconstruction method is the construction of the prior induced on the spectral parameter space by the mapping from the ``physical'' prior on the thermodynamic parameter space.  
The other is the construction of the inverse mapping. 

To illustrate and test the method, we consider two thermodynamic fiducial models: a deflagration and a detonation, each with signal-to-noise ratio around $40$. For each fiducial model we perform 3 MCMC runs: the first samples on the spectral parameters with uniform priors, the second samples on the spectral parameters with an induced prior that is informed by the thermodynamic parameter space, and the last one direct samples on the thermodynamic parameters. For the MCMC runs sampling on the spectral parameters with the induced priors, using our reconstruction method, we also constructed a derived chain of reconstructed thermodynamic parameters.

The success of the method can be judged by its ability to recover the physical parameters and the spectrum of the injected SGWB to $95\%$ confidence. For example, for the deflagration model with $\vw =0.55$, $\al= 0.4$, $r_* = 0.1$, $\log_{10} (\Tn/ \rm GeV) =2.079$ the best constrained thermodynamic parameters are, as can be seen in Table~\ref{table:thermo_deflagration_bounds}, the wall speed $\vw = 0.630^{+0.17}_{-0.059}$ and the nucleation temperature $\log_{10}(\Tn/ \rm GeV) =2.03^{+0.27}_{-0.54}$. The corresponding reconstructed thermodynamic parameters are  $\vw =0.646^{+0.098}_{-0.075}$ and $\log_{10}(\Tn/ \rm GeV)=2.15^{+0.14}_{-0.36}$.  In general the reconstruction method successfully reconstructed the shape of the 1D posterior distributions. The reconstruction could be further improved with a finer grid in the space of thermodynamic parameters used to generate the fit array.

Finally, we highlight the reconstruction method presented here is easily adaptable to different likelihood models (e.g. one with the astrophysical foregrounds included)  without the need to recalculate the physical set of SGWBs. More importantly, in the likely scenario that LISA will release a set of posteriors on generic spectroscopic templates, this method would allow us to extract sound constraints on physical parameters.

\section*{Acknowledgements}
 C.G. is supported by a STFC Studentship. MH (ORCID ID 0000-0002-9307-437X) acknowledges support from the Academy of Finland (grant number 333609). DCH (ORCID ID 0000-0003-2811-0917) acknowledges support from the Academy of Finland (grant numbers 328958 and 353131).

\bibliography{MCMC_SSM}

\end{document}